# Bridging Farm Economics and Landscape Ecology for Global Sustainability through Hierarchical and Bayesian Optimization

Kevin Bradley Dsouza[1*], Graham Alexander Watt[2], Yuri Leonenko[1, 3], and Juan Moreno-Cruz[4]

1 - Department of Earth and Environmental Sciences, University of Waterloo, Waterloo, Canada.
3 - Royal Bank of Canada, Toronto, Canada
2 - Department of Geography and Environmental Management, University of Waterloo, Waterloo, Canada.
4 - School of Environment, Enterprise and Development, University of Waterloo, Waterloo, Canada.
\* - Correspondence: kevin.dsouza@uwaterloo.ca

## Abstract

Agricultural landscapes face the dual challenge of sustaining food production while reversing biodiversity loss. Agri-environmental policies often fall short of delivering ecological functions such as landscape connectivity, in part due to a persistent disconnect between farm-level economic decisions and landscape-scale spatial planning. We introduce a novel hierarchical optimization framework that bridges this gap. First, an Ecological Intensification (EI) model determines the economically optimal allocation of land to margin and habitat interventions at the individual farm level. These farm-specific intervention levels are then passed to an Ecological Connectivity (EC) model, which spatially arranges them across the landscape to maximize connectivity while preserving farm-level profitability. Finally, we introduce a Bayesian Optimization (BO) approach that translates these spatial outcomes into simple, cost effective, and scalable policy instruments, such as subsidies and eco-premiums, using non-spatial, farm-level policy parameters. Applying the framework to a Canadian agricultural landscape, we demonstrate how it enhances connectivity under real-world economic constraints. Our approach provides a globally relevant tool for aligning farm incentives with biodiversity goals, advancing the development of agri-environmental policies that are economically viable and ecologically effective.

## Introduction

Agricultural landscapes cover nearly one-third of global terrestrial lands and are essential to global food security and rural economies [1]. However, modern agricultural practices often prioritize maximizing yield through intensification, characterized by high inputs of synthetic fertilizers, pesticides, irrigation, and specialized monocultures [2]. While these approaches increase productivity, they also impose substantial ecological costs. Agriculture is now a leading driver of global environmental change, contributing to biodiversity loss, habitat fragmentation, soil degradation, water overuse, and greenhouse gas emissions [2, 3]. These impacts increasingly threaten the resilience of agricultural systems themselves, which depend on ecosystem services such as pollination, pest regulation, and nutrient cycling [4].

Agricultural landscapes face a critical dual challenge: increasing food production to meet growing global demand while halting biodiversity decline and restoring degraded ecosystems [29]. This tension emerges from scale mismatches: farmers make decisions based on economic incentives at the plot level, whereas ecological functions like connectivity unfold at the landscape scale [30-35]. Aligning these objectives requires frameworks that bridge farm-level decision-making with regional ecological planning. Although policy instruments such as agri-environmental subsidies aim to bridge the economic–ecological divide, their success in achieving landscape-scale biodiversity outcomes has been mixed [1, 36]. Limitations include poor spatial targeting, fragmented implementation across farm boundaries, and a misalignment between the scale of incentives and the scale at which ecological processes operate [35]. These constraints point to a critical gap: the absence of operational frameworks that can integrate farm-scale economic optimization with landscape-scale ecological planning [37].

Two complementary strategies have emerged as alternatives to conventional intensification. Ecological intensification (EI) aims to enhance ecosystem services while reducing reliance on external inputs like synthetic chemicals [5, 6]. Key strategies include promoting on-farm biodiversity and managing ecosystem functions, particularly insect pollination [5, 7-10] and natural pest control by predators and parasitoids [6, 11-14]. Habitat enhancement practices, such as buffer strips, hedgerows, or rewilded patches, support these functions and can be adapted to a wide range of farming systems [3, 8, 12]. EI is also conceptually compatible with both land-sharing and land-sparing approaches, applied at different spatial scales [8, 15]. Adoption can proceed incrementally, beginning with optimizing land areas for ecological management. While insect population models are often used to assess outcomes [16-19], simpler proxies based on distance and time functions are useful for operationalizing EI at scale [7].

Ecological Connectivity (EC) addresses ecological flows across the broader landscape. EC describes the extent to which landscape structure enables or restricts movement of organisms, genes, and resources among habitat patches [20, 21]. Structural connectivity relates to the physical arrangement of these patches, whereas functional connectivity considers the movement potential for specific species or ecological processes [22-24]. Various tools have been developed to quantify EC, including resistance surfaces, graph theory, and circuit models that represent movement as flows through networks [24-28]. EC is essential for species persistence, genetic exchange, and adaptation, yet is often disrupted by fragmentation in intensively managed agricultural landscapes [20]. Enhancing connectivity is therefore critical to maintaining ecological resilience in agricultural systems.

Existing integrated models typically approach land-use planning by either simulating emergent patterns from farm-level decisions or optimizing land allocation based on multiple objectives [31, 38, 39]. However, few models explicitly optimize the spatial configuration of farm-level interventions, once economically determined, with the sole aim of maximizing ecological connectivity while preserving economic feasibility. Approaches such as the Agroecological Connectivity Index [22] can diagnose connectivity but do not provide spatial optimization. This reveals a methodological gap: how to spatially orchestrate interventions to enhance connectivity without compromising profitability. While spatial optimization can reveal ecologically optimal intervention patterns, implementing such plans is often infeasible due to administrative complexity and coordination burdens among farmers [40, 41]. As a practical alternative, policies can incentivize farm-level decisions through tools like habitat subsidies, eco-premiums, or minimum set-aside mandates [36, 42]. The challenge is to design such instruments so they induce desirable landscape-scale outcomes, like connectivity, without requiring centralized control or coordination.

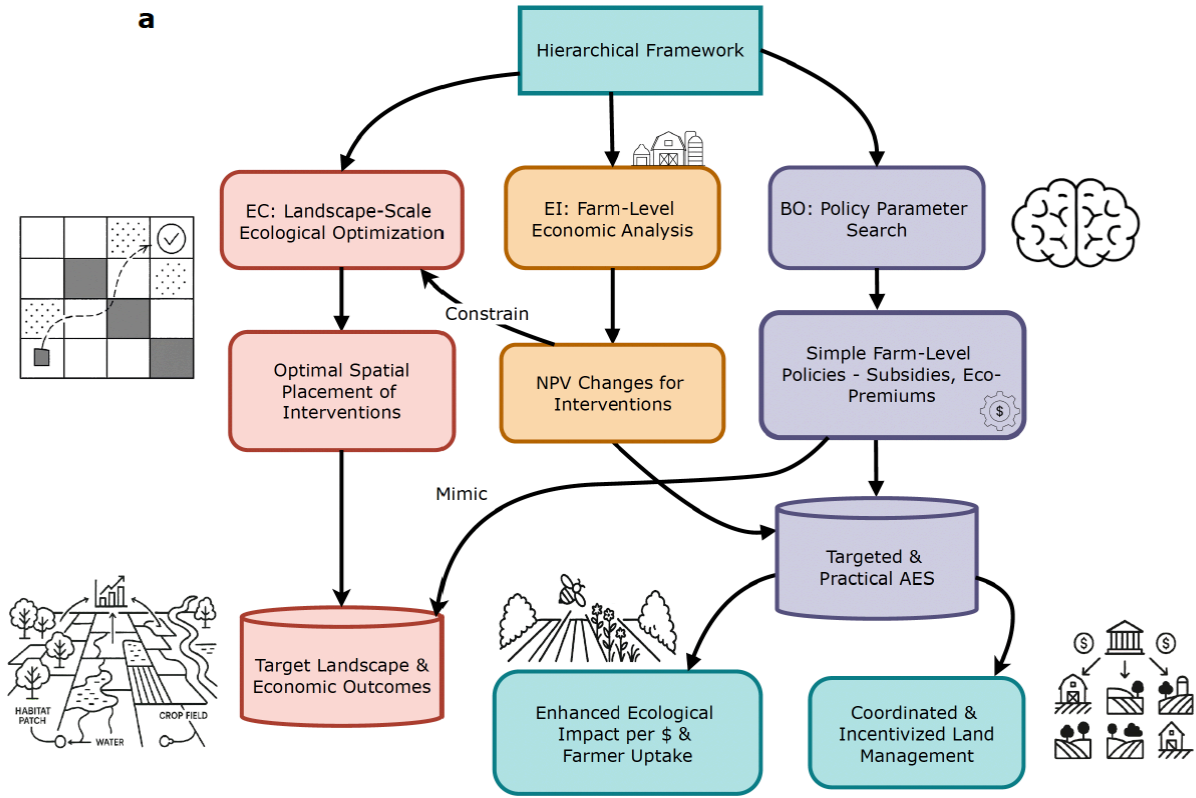

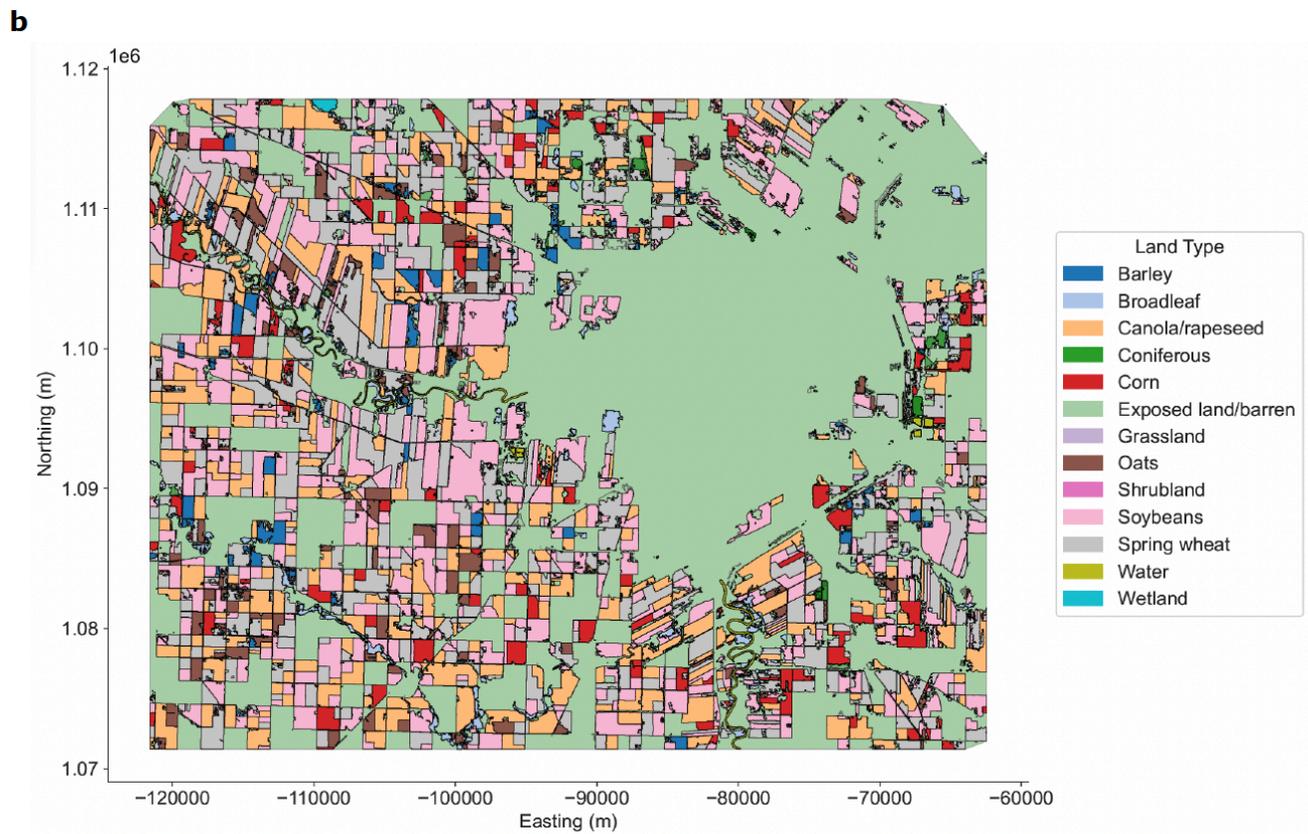

**Fig. 1: The hierarchical optimization framework and demonstration region. a)** Graphical abstract of the three-layer hierarchical optimization framework. The framework sequentially links models across scales: (1) a farm-scale Economic Intensification (EI) model determines economically optimal fractions for interventions; (2) a landscape-scale Ecological Connectivity (EC) model uses these fractions as constraints to optimize the spatial configuration for maximum connectivity; and (3) a Bayesian optimization layer finds simple policy parameters (e.g., subsidies) to replicate the spatially optimized outcomes. This process results in targeted interventions that can increase farmer adoption by being both ecologically effective and economically practical. **b)** The agricultural landscape in Manitoba, Canada, selected as the study area to demonstrate the applicability of the proposed framework.

We introduce a novel hierarchical optimization framework that addresses this integration challenge by sequentially linking farm-level and landscape-level decisions (Fig. 1a). First, a farm-scale Ecological Intensification (EI) model identifies the economically optimal fractions of land to allocate to habitat or margin interventions. These values serve as fixed constraints in a landscape-scale Ecological Connectivity (EC) model, which then determines the spatial configuration of interventions that maximizes connectivity without reducing farm profitability. To support implementable policy, we add a third layer: Bayesian optimization is used to search for simple farm-level policy parameters, such as subsidies or mandates, that replicate the outcomes of the spatial EC optimization (Fig. 1a). This approach enables cost-effective, scalable policy design that aligns farm incentives with biodiversity goals, without requiring spatial coordination.

We demonstrate the full framework in an agricultural region in Manitoba, Canada, highlighting its ability to generate spatially targeted interventions and identify policy instruments that approximate those outcomes under real-world constraints. While applied here in a Canadian context, the modular design of the framework makes it readily adaptable across diverse agricultural systems, from industrialized landscapes governed by the EU's Common Agricultural Policy to mosaic smallholder systems in Latin America and Southeast Asia. Together, the three stages offer a blueprint for designing agri-environmental policies that are economically viable, ecologically effective, and administratively feasible.

# Results

The hierarchical optimization model was applied to real-world agricultural landscapes in a selected region in Manitoba (see Fig. 1b, "Methods" section for more details) to evaluate its performance and identify key drivers for implementing margin and habitat interventions under actual field conditions.

**Farm-Level Ecological Intensification Optimization**

The first stage of our hierarchical framework employs an Ecological Intensification (EI) optimization model operating at the individual farm level. This stage addresses the farm-scale economic considerations critical for farmer adoption. The primary objective is to determine the economically optimal level of investment in two key EI interventions, field margin enhancements and habitat conversion, for each agricultural plot within the farm. Specifically, the model maximizes the total Net Present Value (NPV) of the farm over a defined time horizon $T$, considering a discount rate $r$. The objective function is formulated as:

$$\max \sum_{i \in I_{ag}} NPV_i + \sum_{i \in I_{hab}} NPV_i - \lambda \sum_{i \in I_{ag}} (m_i^2 + h_i^2)$$

where $I_{ag}$ is the set of agricultural plots and $I_{hab}$ is the set of existing habitat plots within the farm. $NPV_i$ represents the Net Present Value calculated for plot $i$, incorporating costs, revenues, and the ecosystem service benefits (pollination and pest control) generated by interventions (see Methods). The decision variables for each agricultural plot $i$ are: $m_i$, the fraction of plot $i$'s area allocated to enhanced field margins ($0 \leq m_i \leq 1$), and $h_i$, the fraction of plot $i$'s area converted to habitat ($0 \leq h_i \leq 1$). The final term is a penalty function with coefficient $\lambda$ applied to the squared decision variables, primarily aiding numerical stability and acting as a form of regularization.

The key outputs of this farm-level optimization are the continuous fractional values $m_i$ and $h_i$ for every agricultural plot $i$. These values represent the economically optimal proportion of each plot to dedicate to margin and habitat interventions, respectively, based on their contribution to the farm's overall profitability via enhanced yields and associated costs. These optimal fractions serve as inputs for the second stage of our framework, which focuses on optimizing the spatial configuration of these intervention amounts to maximize landscape-level Ecological Connectivity (EC). See more details in the "Ecological Intensification (EI) Model" section of the Methods section.

We find that optimal intervention strategies are crop-dependent and differ between intervention types (Fig. 2a). Canola/rapeseed plots distinctly favored margin interventions, showing the highest mean allocation by a large margin. Habitat conversion, while generally implemented at lower levels than margins, was most prominent in Soybean (mean≈0.055) and Corn (mean≈0.05) plots (Fig. 2a). Other crops like Barley, Oats, and Spring Wheat received minimal allocations of either intervention type. This divergence reflects the interplay of parameterized benefits, costs, and crop prices. Higher intervention adoption in Canola and Soybeans is driven by their substantial associated ecosystem service benefits ($\alpha, \delta$, etc.) and/or high market value ($p_c$), which justify intervention costs (see "Methods" section for more details). Economic factors, particularly maintenance costs, significantly influenced the overall intervention strategy chosen. Simulations resulting in habitat conversion ('Habitat' states) were associated with higher ratios of agricultural maintenance cost relative to habitat maintenance cost (Fig. 2b). Correspondingly, the decision to implement margins ('Margin' state) was linked to lower ratios of margin maintenance cost relative to habitat maintenance cost (supplementary Fig. 3a). Correlation analysis weakly confirmed these trends, showing habitat conversion levels were positively driven by higher agricultural maintenance costs (r=+0.03) and negatively by higher habitat maintenance costs (r=−0.03) (Fig. 2c). Among other factors, Canola/rapeseed price affected both interventions the highest (r=+0.05), because of its strong ecosystem service benefits (Fig. 2c, supplementary Fig. 3c).

Optimal fractions of EI interventions for a selected configuration of farms (supplementary Fig. 4a, b) is shown in Fig. 2d. These are computed by the EI optimization model, which calculates the most economically beneficial fraction of interventions, considering factors like distance-based pollination, pest control, and crop yield (supplementary Fig. 6,7, see "Methods" section for more details). Habitat conversions are very rare and margin interventions are largely preferred (Fig. 2d). The extent of interventions are dependent on the value of the regularizing factor ($\lambda$) and the distance threshold for intervention effects in the EI optimization (see supplementary Fig. 5, "Methods" section for details). We also investigated the sensitivity of the EI model's outputs to variations in crop-specific ecological parameters. Specifically, we examined how changes in the parameters governing pollination (alpha, beta, gamma) and pest control (delta, epsilon, zeta) benefits derived from margins and habitats for individual crops influence the average intervention fractions (supplementary Fig. 6,7). A consistent finding across both intervention types is that the majority of crop-specific ecological parameters exhibit Pearson correlations very close to zero with the respective mean allocation fractions (supplementary Fig. 8,9). This indicates that, within the tested range of variation (0.5x to 2.0x) and under simultaneous multi-parameter changes, average allocation decisions for both margins and habitat conversions are not strongly driven by linear changes in any single ecological parameter for a specific crop.

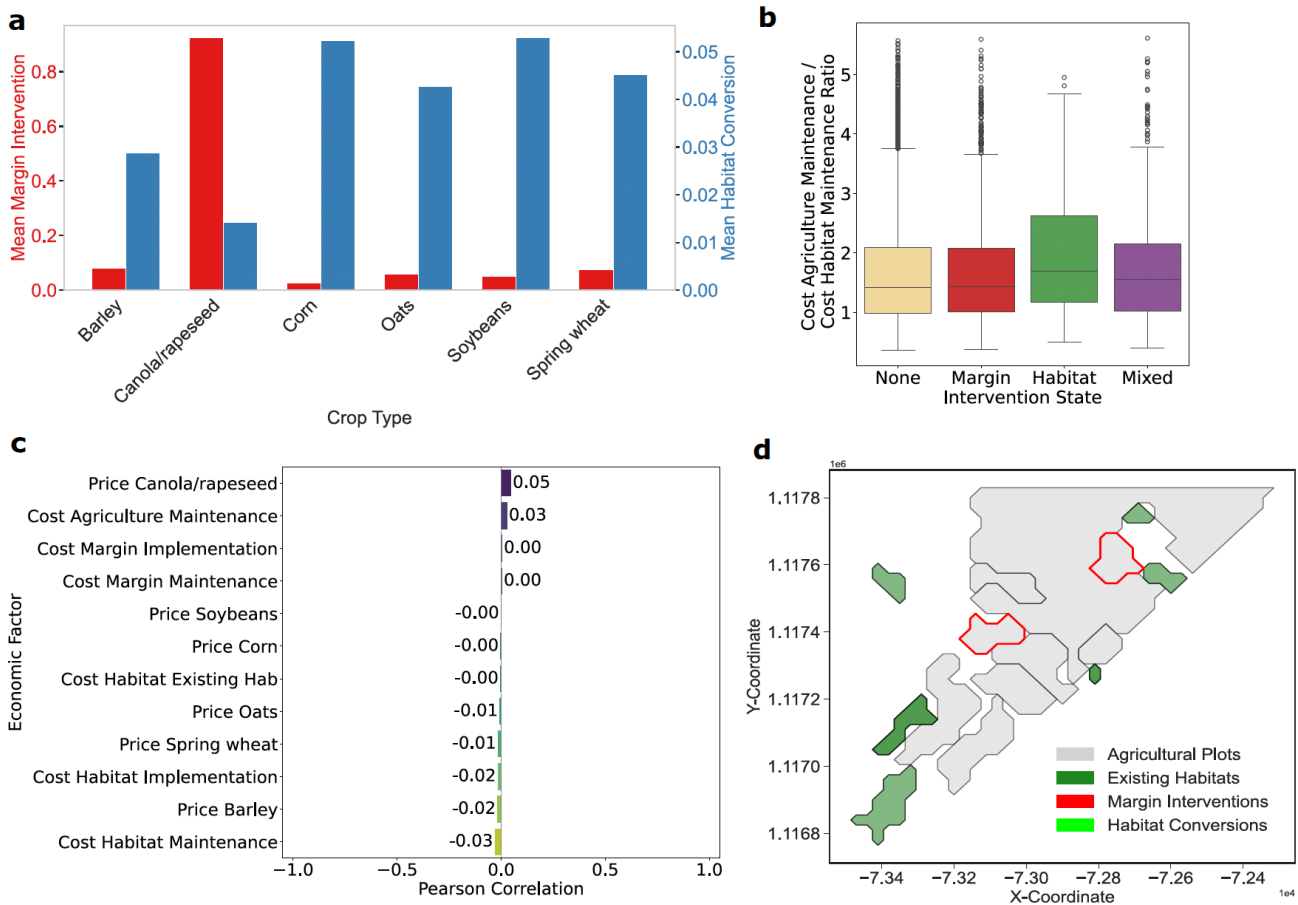

**Fig. 2: Results from the Economic Intensification (EI) model. a)** Optimal fractions of margin and habitat interventions vary by crop type. Margin interventions are heavily favored in high-value Canola/rapeseed plots, while habitat conversion is prominent, at lower levels, in Soybean and Corn. **b)** A box plot of intervention states as a function of the ratio of agricultural maintenance and habitat maintenance costs. Habitat conversion is linked to a high ratio of agricultural maintenance cost relative to habitat maintenance cost. **c)** Sensitivity analysis illustrating how economic factors, including crop prices and maintenance costs, influence the level of habitat conversion. **d)** An example of an optimal intervention layout determined by the EI model for a selected farm configuration, highlighting the general preference for margin interventions over habitat conversions.

The first stage of our framework reveals that economically optimal ecological interventions vary substantially by crop type and intervention category. Margin enhancements were most prominent in high-value crops like canola, which benefit significantly from ecosystem services, while habitat conversion was limited overall but concentrated in soybean and corn plots. The model also highlights how profitability considerations, shaped by crop price, yield, and maintenance costs, strongly influence intervention adoption. Importantly, sensitivity analyses show that average intervention levels are relatively robust to variation in individual ecological parameters, suggesting that the model's outputs are driven more by economic than ecological uncertainties. Overall, our findings show that crop- and context-specific economic calibration needs to be carried out when designing ecological intensification strategies.

## Landscape Connectivity Enhancement via EC Optimization

Building upon the farm-level optimization of intervention extents determined by the EI model, the subsequent landscape-level Ecological Connectivity (EC) model addresses the challenge of spatially configuring these interventions. This second stage of our hierarchical framework aims to maximize a chosen landscape connectivity metric across a given farm configuration to best facilitate ecological flows and species movement.. The EC model determines the optimal placement of field margins and habitat patches, the amounts of which are guided by the economic outcomes of the preceding EI stage.

A core feature of this EC optimization is its adherence to the economic realities established at the farm level. The model ensures that the reconfigured landscape, while optimized for connectivity, maintains the economic performance for each farm. This is achieved by constraining the recalculated Net Present Value of each farm ($N_{f,new}$) to remain above a specified threshold relative to the baseline Net Present Value ($N_{f,base}$) established from the EI stage's outputs, governed by a maximum allowable loss ratio ($\rho_{max}$).

$$N_{f,new} \geq (1 - \rho_{max})N_{f,base}$$

To achieve this, agricultural plots are discretized into candidate linear arcs for margins and polygonal cells for habitat. The EC model employs decision variables to select the optimal combination of these candidates, which is driven by the objective of maximizing landscape connectivity using the Integral Index of Connectivity (IIC). The final output is a spatially explicit plan detailing the locations of ecological interventions that enhance connectivity without compromising the predetermined farm-level economic viability. The detailed mathematical formulation of this EC model is presented in the "Ecological Connectivity (EC) Model" section of the Methods and see "Objective Function" in this section for details on how IIC is defined.

We analyzed the effectiveness of two distinct optimization stages, repositioning and connectivity optimization, across the simulated farm landscape configurations. The repositioning stage aimed to maximize connectivity by rearranging pre-defined fractions of margin and habitat interventions within each agricultural plot, without altering the total amount of intervention per plot. The connectivity optimization stage allowed for the selection of new or reconfigured margin and habitat pieces across the entire landscape, aiming to maximize connectivity while ensuring that the Net Present Value (NPV) for each farm did not decrease by more than a specified threshold ($\rho_{max}$). The Integral Index of Connectivity (IIC) was used as the quantitative metric for landscape connectivity. The connectivity optimization yields substantially higher IIC scores compared to repositioning (Fig. 3a), with the mean IIC score increasing from $6.3 \times 10^3$ after repositioning to $7.9 \times 10^3$ after optimization, representing a considerable improvement in landscape connectivity. The entire distribution of connectivity scores shifted towards higher values after the optimization. While repositioning can offer some improvement by optimizing spatial arrangement, allowing the optimization framework to select both the location and the amount (subject to economic viability) of interventions provides far greater potential for enhancing landscape-scale ecological connectivity (Fig. 3a).

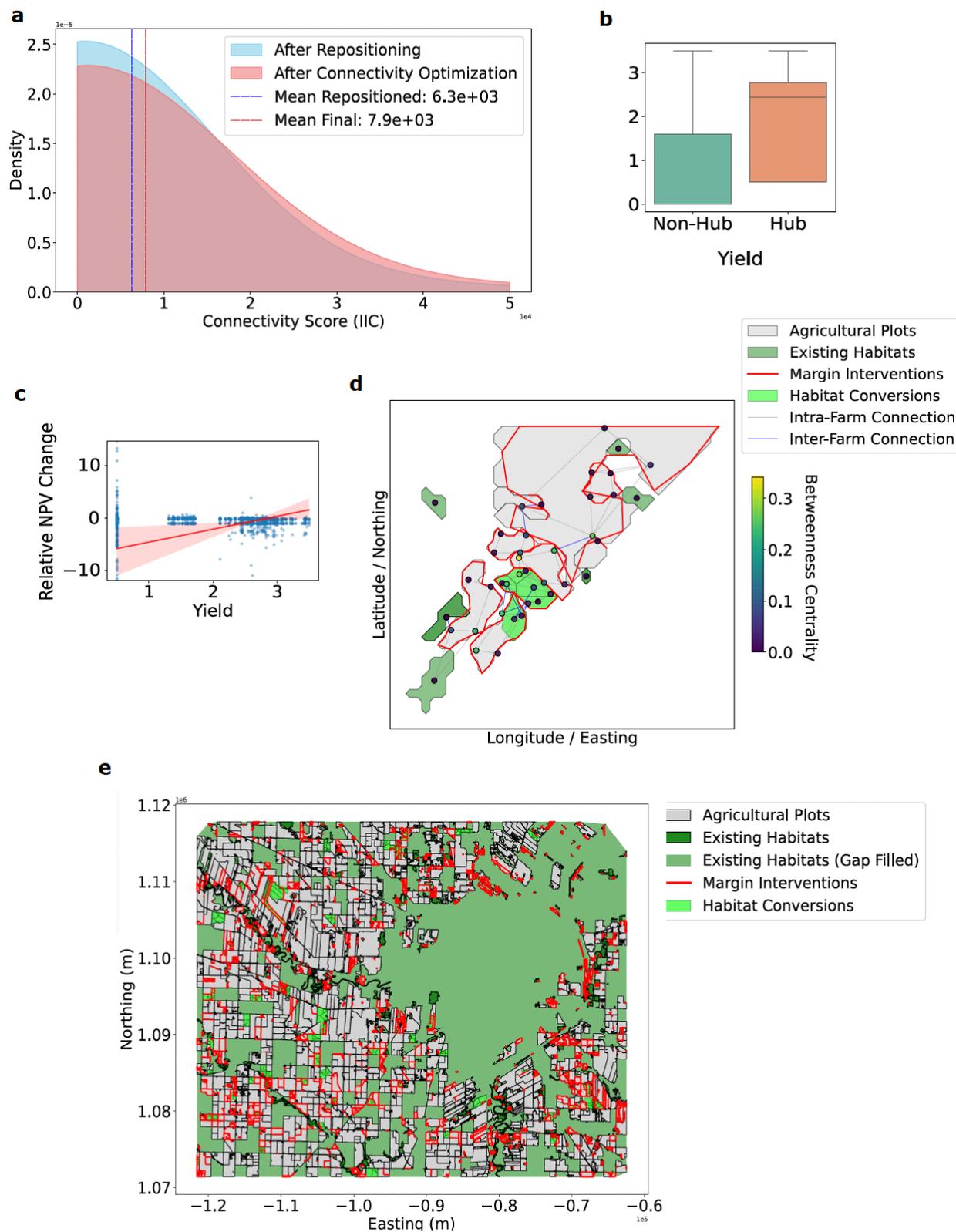

**Fig. 3: Results of the Ecological Connectivity (EC) optimization model. a)** The distribution of landscape connectivity scores (IIC) obtained from the configurations after each optimization stage. The full "Connectivity Optimization" stage (red)

results in a substantial improvement in landscape connectivity compared to the simpler "Repositioning" of interventions (blue). **b)** Box plot of aggregate yield for hub and non-hub plots. Hub plots exhibit a higher median yield, indicating a greater economic capacity to sustain ecological interventions. **c)** Scatter plot of relative change in NPV after connectivity optimization compared to EI baseline versus baseline yield. Plots with higher intrinsic yields tend to experience smaller relative NPV reductions. **d)** Optimal interventions decided by the EC model for a selected configuration of farms, along with plot level betweenness centrality and intra-farm (dashed gray)/inter-farm (solid blue) connections. Plot-level betweenness centrality highlights the importance of certain plots as stepping stones in the network. **e)** Composite map showing the spatial distribution of all optimal interventions selected by the EC model across the landscape, revealing the most critical areas for landscape-scale connectivity enhancement.

To explore the connectivity-optimized structure further, we analyzed the resulting configuration of habitat patches and field margins. This involved identifying plots acting as critical connectivity hubs, characterizing the structure of the enhanced habitat network, and examining the distinct roles plots play in connectivity. The optimized spatial arrangement of selected interventions was represented as a spatial graph, with nodes corresponding to interventions and edges representing adjacency. Betweenness centrality was calculated for each node to quantify its importance as a "stepping stone". These centrality scores were aggregated (summed) to the level of the original plots. Agricultural plots ranking within the top 10% by aggregated betweenness centrality were designated as "connectivity hubs". We then conducted a comparative analysis across all configurations to determine whether designated hub plots exhibit distinct characteristics compared to non-hub agricultural plots (Fig. 3b).

The aggregate analysis reveals notable trends: hub plots tend to have higher median yields (Fig. 3b), more complex shapes (higher perimeter, higher area, more sides), and more neighbors (supplementary Fig. 10). Hub plots have higher median base yields potentially because plots with initially higher base yields are better able to economically sustain the implementation of connectivity-enhancing interventions. The yield boost from ecosystem services on these high-yield plots helps to ensure that the farm's overall profitability (NPV) remains within acceptable limits while the interventions contribute to the landscape connectivity goals. The model favors solutions that are both ecologically effective (high connectivity) and economically viable (such as margin interventions that don't take away agricultural land). High base yield plots provide more flexibility to implement these ecological interventions. On the other hand, the geometric distinctiveness of hubs contributes by providing more extensive interfaces for connecting with other habitat patches and margins.

According to our EC model, as farms are permitted to potentially absorb a larger reduction in NPV ($\rho_{max}$), the optimization model can configure interventions to yield higher IIC values. To understand the distribution and drivers of these economic impacts at a finer scale, we analyzed the results aggregated across multiple landscape configurations, by determining the relative change in NPV resulting from the connectivity-enhancing interventions. Baseline yield exhibits a positive correlation with the relative NPV change (Fig. 3c). The regression line suggests that plots with higher intrinsic yields tend to suffer smaller relative NPV losses. This could be attributed to the optimization process, while maximizing landscape-level connectivity, implicitly favoring interventions in areas where the opportunity cost (foregone yield from the highest-productivity land) is lower, thus relatively shielding the highest-yielding plots from negative impacts.

Next, we visualized the resulting network structure, including interventions, centrality scores, and connections (Fig. 3d). We saw that high centrality nodes (yellow circles) are visibly positioned in locations critical for network cohesion, often bridging gaps or linking different habitat clusters via both intra-farm (dashed gray) and inter-farm (solid blue) connections (Fig. 3d). These findings demonstrate the critical role of hub plots. Their larger size and complex shapes (supplementary Fig. 10), appear well-suited for this bridging role, acting as efficient stepping stones and habitat anchors. The optimization framework effectively identifies and leverages plots whose specific location and configuration provide high value for landscape-scale cohesion by facilitating links between farms. This demonstrates that effective landscape connectivity planning in fragmented agricultural systems likely requires strategies targeting the establishment or enhancement of habitat links across ownership or management boundaries. A comparison between the initial EI outcomes (Fig. 2d) and the connectivity optimized outcome (Fig. 3d,e) demonstrates the effectiveness of the connectivity-driven optimization. In contrast to EI outcomes, the optimized layout displays a denser network of interventions. The optimization selects additional margin segments and introduces new habitat patches within agricultural plots. The placement of these interventions in the optimized landscape is visibly strategic and non-random: margins often delineate boundaries or connect to existing habitat blocks, while habitat patches are frequently positioned adjacent to these margins or existing habitats, forming clusters, bridging gaps, and creating more continuous corridors or stepping stones (Fig. 3d,e).

Overall, the second stage demonstrates that spatial optimization of interventions significantly enhances landscape connectivity without compromising farm-level economic viability. Compared to simple repositioning, the EC optimization yields notably higher connectivity scores by strategically clustering habitat and margin interventions to form continuous corridors. High-yield and geometrically complex plots emerged as critical connectivity hubs, offering both ecological leverage and economic buffer capacity. These findings show that connectivity goals can be advanced by prioritizing plots with high ecological centrality and economic headroom, offering a blueprint for spatial targeting under real-world constraints

## Bayesian Optimization Finds Policy Instruments within EI to Mimic EC Outcomes

We employ Bayesian Optimization [43] to find the parameters ($\theta_P$) of non-spatially explicit, farm-level EI policy instruments. These instruments include subsidies for habitat creation and maintenance, payments per hectare for habitat, mandates for minimum ecological set-asides, and eco-premiums for sustainably grown crops (for a full list of policy instruments and their search ranges see Table 3 in Methods). The BO systematically searches for policy parameter values that, when implemented, incentivize farmers to adopt ecological interventions. The goal is to achieve landscape connectivity ($C_L^{policy}$) and farm economic outcomes ($NPV_f^{policy}$) that closely match the targets ($C_L^{target}, NPV_f^{target}$) established by the direct EC optimization in the second stage. The objective function for the BO, $O_{BO}$, aims to minimize a weighted sum of the differences between the policy-induced outcomes and the target outcomes, averaged over multiple landscape configurations ($k$):

$$O_{BO}(\theta_P) = \frac{1}{N_S} \sum_{k=1}^{N_S} \left( w_C \cdot |C_{L,k}^{policy}(\theta_P) - C_{L,k}^{target}| + w_{NPV} \cdot \text{mean}_{f \in F_k}(|NPV_{f,k}^{policy}(\theta_P) - NPV_{f,k}^{target}|) \right)$$

This optimization is also constrained by a maximum allowable average policy budget ($B_{max}$). For the current study, this was set to $B_{max} = 500,000$. This value is flexible and can be adjusted to reflect different regional economic constraints, policy priorities, or available public funds. See "Bayesian Optimization (BO) to Find Policy Instruments" in Methods for more details.

The convergence plot (Fig. 4a) illustrates the progress of the optimization. It shows the minimum objective function value (average absolute connectivity difference) found as the number of evaluations increased. Significant improvements occurred within the first ~50 calls, after which the rate of improvement slowed, indicating that the BO effectively explored the parameter space and likely converged towards a near-optimal solution within the defined search space and evaluation limit. We re-evaluated the top 50 policies identified by the BO to understand their performance in terms of average connectivity and average farm NPV (Fig. 4b).

Many BO-derived policies achieve average connectivity scores comparable to, or even exceeding, the "Baseline - Optimized" scenario (Green Star). Crucially, these ecological gains are often accompanied by higher average farm Net Present Values (NPVs) compared to this same optimized baseline. This improvement in average farm profitability under BO policies is largely because the policy evaluations did not impose the same farm-level NPV or connectivity constraints that might have limited the "Optimized Baseline" during its own computation (supplementary Fig. 11a). By providing incentives and altering the economic landscape for farmers, these policies can lead to voluntary adoption of practices that enhance connectivity, potentially overcoming the more restrictive conditions of the baseline optimization. Furthermore, the BO policies consistently outperform the "Baseline - Repositioned" scenario (Blue Star) in terms of connectivity, where farmers act purely on individual NPV optimization without policy guidance or landscape coordination, demonstrating the clear benefit of targeted policy interventions.

The relationship between average policy cost (NPV of government expenditures), reveals that significant ecological improvements do not invariably necessitate high public expenditure (Fig. 4c). For instance, the policy achieving the highest observed connectivity is realized at a moderate-high government cost, far from the most expensive options (Fig. 4c). Conversely, some policies like the one with the lowest observed connectivity, despite incurring substantial costs, yield poor connectivity outcomes, even falling below baseline levels (Fig. 4c). This highlights the critical role of optimization in policy design to ensure cost-effectiveness and avoid inefficient allocation of resources. Many policies elevate connectivity scores above both the "Repositioned" and "Optimized" baseline connectivity levels across a spectrum of costs, providing decision-makers with a portfolio of options that balance ecological targets with budgetary constraints.

Next, we examine key policy parameter distributions in top-performing policies (Fig. 4d). We see that effective designs often involve moderate subsidy levels for intervention establishment and maintenance, with medians for various factors typically ranging from 0.2 to 0.3 within their 0.0-0.5 search spaces. However, the mandate parameter *Min. Margin Frac. Adj. Habitat* (Mandated minimum fraction of margin adjacent to existing habitats, with a 0.0-0.3 search range) consistently skewed towards lower values, with a median around 0.1 in these successful policies (Fig. 4d). For the full set of parameter distributions see supplementary Fig. 12. Exploring this further, we visualize specific successful policy archetypes, including parameters like eco-premiums, habitat area payments, and total habitat mandates. These profiles illustrate how different optimization objectives shape the policy levers. For example, a policy targeting good connectivity with maximum NPV and minimal deviation from baseline connectivity tends to employ substantial eco-premiums across multiple crops and moderate habitat maintenance subsidies (Fig. 4e).

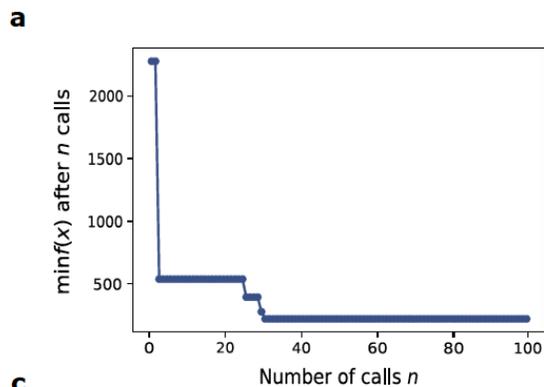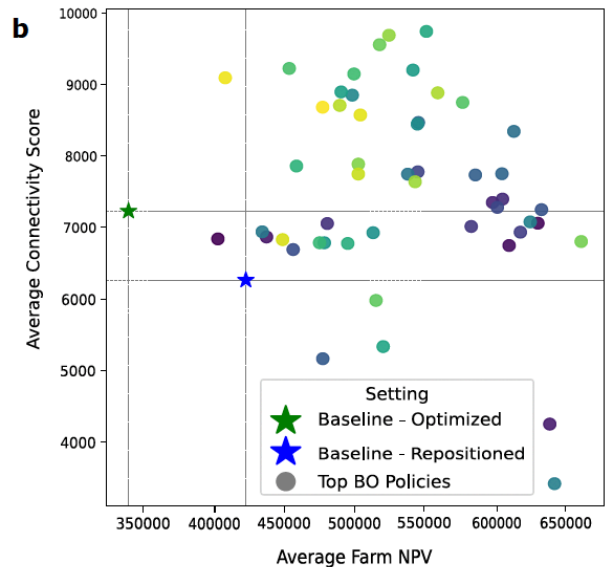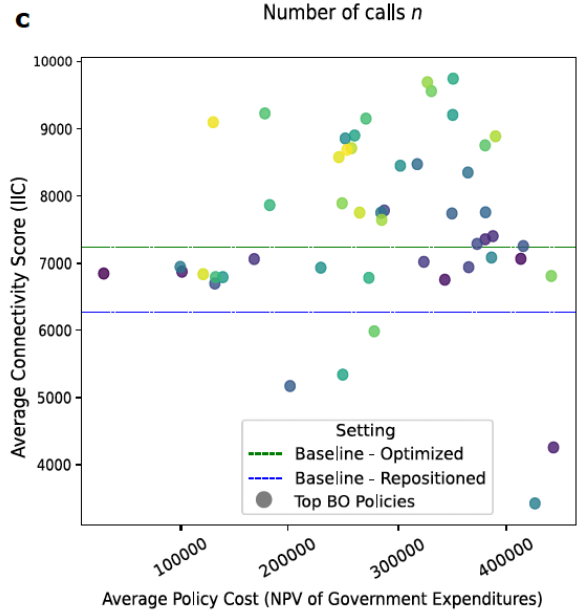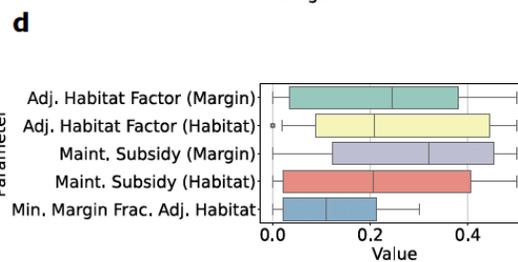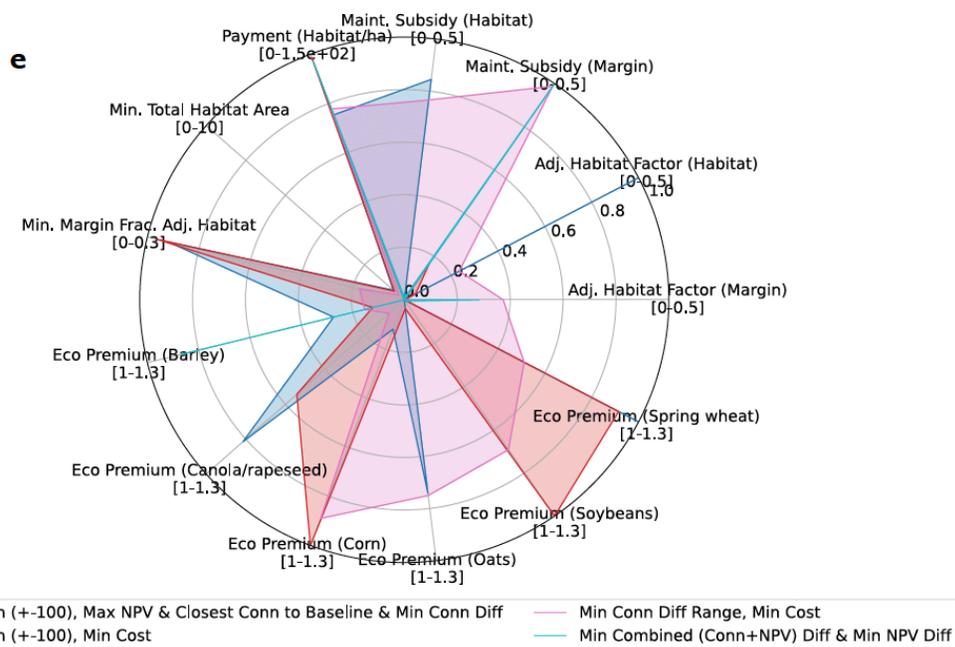

**Fig. 4: Bayesian Optimization (BO) Results Identify a Portfolio of High-Performing Agri-Environmental Policies.** This figure provides a summary of the results from the BO process, from search dynamics to the characteristics of effective policies. **a)** Minimum objective function value found with increasing number of evaluations. The search process is efficient, which rapidly identifies near-optimal solutions within approximately 50 evaluations. **b)** Average connectivity vs average farm NPV for top 50 policies identified by BO, along with baseline optimized and repositioned outcomes. Well-designed incentive policies outperform baseline strategies. **c)** Average connectivity vs average policy cost for top 50 policies identified by BO. Ecological gains do not invariably require maximum public expenditure; instead, optimization is key to designing cost-effective interventions and avoiding inefficient spending. **d)** Policy parameter distributions in top-performing policies. There is a preference for moderate subsidy levels and relatively low mandatory habitat adjacency requirements, suggesting gentle nudges can be highly effective. **e)** Radar profiles of specific successful policy archetypes, including closest connectivity to baseline, minimum cost, maximum farm NPV. etc. Different strategies, from incentive-heavy approaches using eco-premiums to more frugal, cost-conscious designs, can achieve desirable outcomes.

In contrast, policies prioritizing minimal cost while achieving good connectivity generally exhibit lower subsidy rates and eco-premiums. These results highlight that effective policy design can involve diverse strategies, from incentive-heavy approaches utilizing eco-premiums and subsidies, to those more reliant on area-based mandates, depending on the specific desired outcomes and constraints. The full list of policy instrument parameters for well performing policy archetypes is shown in Table 1 in supplementary, and a heatmap of effects of all the policy dimensions on the key metrics are presented in supplementary Fig. 13.

To summarise, using Bayesian Optimization, we identified a range of cost-effective policy instruments, such as eco-premiums, habitat payments, and minimum margin mandates, that approximate the landscape-scale outcomes of spatial EC optimization. The top-performing policies consistently achieve high connectivity and favorable economic outcomes within realistic public budget constraints. Notably, different optimization goals produce distinct policy archetypes, from incentive-heavy packages to more cost-conscious mandates. This approach provides a scalable method for designing agri-environmental policies that are not only ecologically effective and economically viable, but also administratively feasible, offering policymakers a portfolio of flexible, evidence-based options.

## Potential Applications and Extensions

Our hierarchical EI-EC-BO framework is applicable to a variety of land management challenges that require developing targeted, adaptive, and economically sound strategies that balance agricultural production with environmental stewardship.

For instance, the framework can be used to design effective and implementable agri-environmental schemes (Fig. 1a). This can be accomplished by the EI component evaluating farm-level economic impacts of interventions, and the EC component determining optimal spatial placement for maximum ecological benefit. Importantly, where direct EC optimization might yield complex plans, the BO component drives practical implementation by identifying simpler, non-spatially explicit farm-level policies (like subsidies or eco-premiums) that achieve comparable landscape and economic outcomes. Moreover, the framework's utility extends to integrating multiple ecosystem services. To accommodate this, the EI model's NPV calculation can be broadened to include values or costs associated with services like carbon sequestration from agroforestry or water regulation through wetland restoration. Similarly, the EC stage's objective function can be adjusted to co-optimize for connectivity alongside other spatial goals, such as maximizing carbon storage hotspots or minimizing downstream flood risk, thereby enabling the design of interventions that yield multiple co-benefits. The BO component can then be utilized to find the most effective and efficient policy combinations, for instance, mixes of carbon payments, water quality incentives, and biodiversity subsidies, that encourage land management practices delivering these combined benefits, as identified by the EI/EC stages.

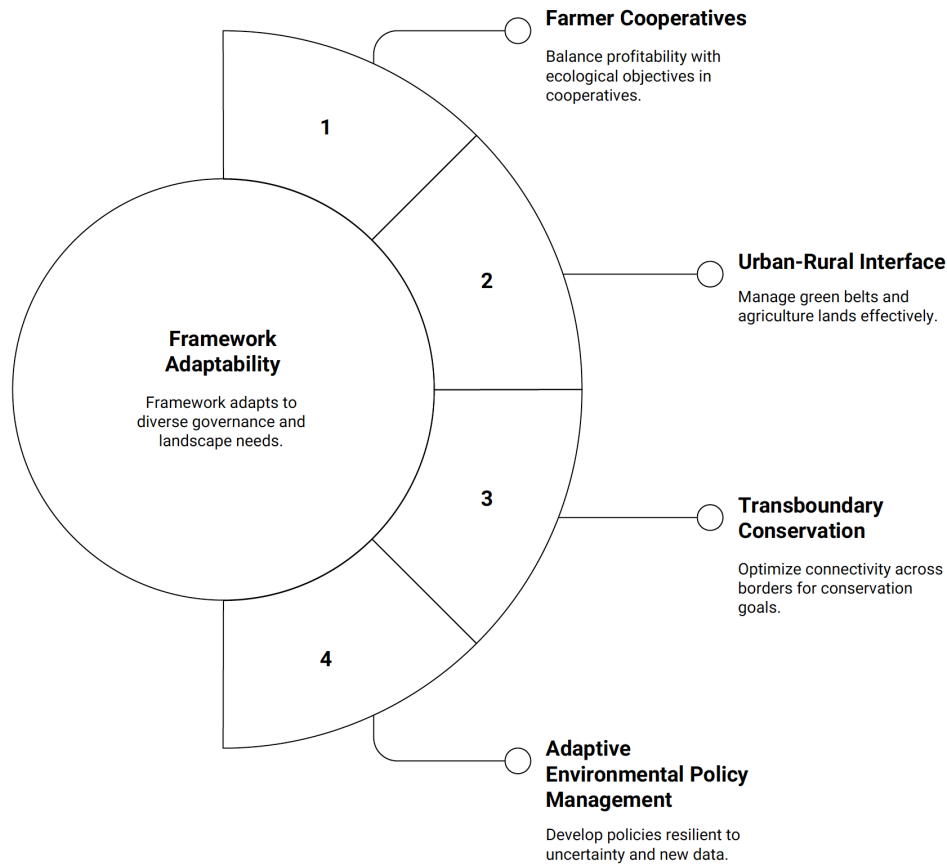

**Fig. 5: Applications and extensions of the proposed framework.** This figure illustrates the versatility of the framework for designing tailored and adaptive environmental policies across diverse and complex scenarios. The model's components can be adapted for various governance structures, such as helping farmer cooperatives align individual member profitability with collective ecological goals, or aiding planners in the urban-rural interface to design incentives that preserve green belts under development pressure. For larger-scale challenges like transboundary conservation, the framework can be used to harmonize policies across different jurisdictions to meet shared environmental targets. A crucial, cross-cutting application is its inherent support for adaptive management.

Our framework's adaptability facilitates tailored policy design and adaptive management across diverse governance structures and landscapes (Fig. 5). Farmer cooperatives, for example, could use the EI/EC components to balance individual member profitability with collective ecological objectives; BO can assist these cooperatives in designing internal incentive mechanisms or in advocating for external policies that align individual actions with these group aims. In the urban-rural interface, planners might apply EI/EC logic to manage green belts or community agriculture lands, with BO helping to design incentive programs that encourage land uses, maintaining connectivity and ecosystem services under development pressures. For transboundary conservation, the EI model can be applied reflecting each jurisdiction's specific economic conditions, while a coordinated EC stage optimizes connectivity across borders based on shared ecological goals. BO can then aid in crafting harmonized or complementary policy instruments across these jurisdictions. Crucially, the BO process allows for the development of policies robust to uncertainty by evaluating their performance across multiple landscape configurations and budgetary scenarios. This iterative nature also means that as new data on farmer responses or ecological outcomes become available, the BO can be rerun to refine and adapt policy instruments, supporting an adaptive management approach to environmental policy.

Thus our hierarchical framework, by integrating economic optimization, spatial ecological planning, and sophisticated policy search, offers a toolkit for addressing the complexities of sustainable land management in agricultural landscapes.

## Discussion

In this study we present a hierarchical optimization framework that effectively advances ecological connectivity in agricultural landscapes while maintaining farm-level economic viability. By decoupling and sequentially linking farm-scale and landscape-scale decisions, the framework first applies Ecological Intensification (EI) to determine economically optimal intervention levels and then leverages these outputs in a landscape-level Ecological Connectivity (EC) model to optimize spatial arrangement. This tiered approach ensures that ecological gains are pursued within economically feasible bounds. The integration of continuous farm-level optimization with discrete spatial planning distinguishes this framework from previous models that either simulate emergent patterns or simultaneously optimize conflicting objectives, often at the expense of connectivity [31, 38, 39]. Importantly, the addition of a Bayesian Optimization (BO) layer enables the identification of simple, non-spatial policy instruments, such as subsidies or eco-premiums, that approximate the benefits of spatially explicit interventions. Together, these elements offer a practical blueprint for designing agri-environmental policies that are both spatially informed and administratively feasible. By operationalizing a layered optimization structure, this framework enables policymakers to design spatially targeted, evidence-based interventions that were previously infeasible due to administrative complexity or lack of integration between economic and ecological planning.

Despite its strengths, the framework has limitations that should inform its interpretation and future development. The ecological modeling relies on simplifications, including exponential decay functions and fixed accumulation parameters, and uses the Integral Index of Connectivity (IIC) as a proxy for functional connectivity. Economic parameters such as crop prices and input costs are held constant over time, which may limit realism under dynamic market or climate conditions. Computational complexity is also a constraint: the EC model's formulation as a Mixed-Integer Nonlinear Program (MINLP) limits its scalability to large, high-resolution landscapes. Additionally, outcomes are sensitive to the geometric complexity of farm plots, more irregular configurations may yield divergent results, as shown in synthetic landscape experiments (see "Synthetic Data Experiments" sections in the Methods and supplementary). While sensitivity analyses addressed some parameter uncertainties, the framework currently does not model farmer heterogeneity or behavioral responses explicitly, though the BO layer partially accounts for economic responsiveness. In addition to technical and ecological uncertainties, implementation will also depend on governance capacity, the administrative burden of managing spatially explicit interventions, and farmer perceptions of fairness and feasibility. These social dynamics are critical to uptake, particularly in smallholder or low-resource contexts.

Future work can extend the framework in several directions to enhance realism, scalability, and policy relevance. Integrating more sophisticated ecological models, including species-specific dispersal behavior, habitat quality gradients, and landscape matrix effects, would improve ecological fidelity [44]. Accounting for temporal dynamics, such as climate change, market variability, or evolving policy environments, is critical for long-term planning. Addressing computational challenges through heuristics or decomposition methods would enable applications to larger or more fragmented landscapes. Empirical validation using real-world farm data and ecological monitoring would increase confidence in predicted outcomes. Finally, coupling the framework with agent-based models of farmer behavior [38, 45, 46] could provide insight into adoption patterns and the social dynamics that influence policy uptake, supporting the design of more equitable and effective incentive schemes.

The hierarchical framework presented here offers a scalable, decision-support tool for reconciling agricultural production with ecological conservation. By integrating farm-level economic optimization with landscape-scale ecological planning, it provides a rare bridge between localized decision-making and global biodiversity goals. Its capacity to identify not just how much intervention is economically viable, but also where it should be located for maximal ecological impact, positions it as a powerful engine for spatially targeted PES schemes and conservation programs. The Bayesian Optimization component strengthens real-world applicability by translating complex spatial strategies into simple, incentive-based policies. Our proposed framework is generalizable, modular, and one that is well-suited to diverse agricultural contexts and policy environments, from the Common Agricultural Policy in Europe to smallholder systems in the Global South. As countries work toward SDGs and the Kunming-Montreal Global Biodiversity Framework, such tools are essential to design evidence-based, cost-effective, and scalable agri-environmental policies. Taken together, the EI, EC, and BO layers form a modular architecture for integrated land-use planning, simultaneously addressing economic viability, ecological integrity, and administrative feasibility, three pillars of sustainable agri-environmental policy. As countries operationalize ecosystem service payments and biodiversity targets, this approach offers a replicable model for spatially aware, cost-effective conservation.

# Methods

### Agricultural Landscape Data
We used the Canadian annual crop inventory (CACI) of 2022 [47], produced by Agriculture and Agri-Food Canada (AAFC). This dataset comprises digital maps identifying crop types and land cover across all Canadian provinces. ACI raster data of a selected region in Manitoba and its corresponding classification legend, which maps integer or RGB values to land cover labels (e.g., "Spring wheat", "Grassland", "Water"), were used as the primary inputs. Initial processing involved converting the relevant raster data segments into vector format. Pixel groups with identical values (representing distinct land cover patches) in the input raster were identified, boundaries of these pixel groups were traced, and these boundaries were subsequently converted into vector polygons using the raster's georeferencing information. Each resulting polygon was assigned a 'label' corresponding to its land cover type, derived by matching its original pixel value to the provided classification legend. This process generated an initial set of land cover polygons stored in a GeoJSON format. Coordinate transformations, such as converting between geographic (latitude/longitude) and projected coordinates (e.g., Albers Equal Area Conic suitable for Canada), were performed as needed during intermediate steps, ensuring consistent spatial referencing and accurate area and distance calculations in meters.

Following the initial vectorization, agricultural plots were grouped into farm units, and associated habitat patches were identified and processed. Polygons representing specific agricultural crop types (Barley, Canola/rapeseed, Corn, Oats, Soybeans, Spring wheat) were identified from the initial plot data. Adjacent crop polygons were iteratively grouped together to form candidate farm units. Grouping aimed to aggregate neighbouring plots, potentially merging smaller adjacent groups, while implicitly limiting excessive aggregation through neighbour search constraints. The final geometry for each farm was generated by dissolving the boundaries between its constituent crop plots. For each defined farm unit, nearby non-crop "habitat" polygons (including Broadleaf, Coniferous, Exposed land/barren, Grassland, Shrubland, Water, Wetland) were identified. Habitat plots located within the farm's boundary or within a 200-meter buffer were considered associated, with a limit of the 5 closest or contained habitat plots being linked to the farm. Adjacent habitat plots of the same label (e.g., two touching 'Grassland' polygons) were merged into single, larger habitat polygons.

Individual farm landscapes were then extracted and refined. For each farm ID, a separate GeoJSON file was created. This file contained the farm's main boundary polygon, all its constituent agricultural plot polygons, and all its associated habitat polygons. Yield information was integrated using a biomass inventory dataset (Biomass Inventory Mapping and Analysis [48]). For each agricultural plot polygon within a farm, the spatially corresponding yield value was retrieved from the biomass inventory. For crops not explicitly mapped, an average yield was calculated from the available mapped crop yield values within the overlapping biomass polygon. Yields were converted to Tonnes per Hectare (t/ha). Agricultural Plots with 0 yield were assigned a minimum yield of 0.5 and habitat plots were assigned a yield of 0. To remove potentially insignificant sliver polygons or artifacts, a size-based filtering step was applied to each farm. Polygons with areas below the 25th percentile of all plot areas within that specific farm were removed. Internal gaps or holes within the remaining polygons of a farm were identified and were filled by creating new polygons. The land cover label assigned to a gap polygon was determined by the majority label of the immediately adjacent polygons. These gap-filling polygons were added to the farm's feature set, creating a spatially complete representation within the farm's boundary.

To analyze broader landscape patterns, individual farms were grouped into spatially coherent configurations. An adjacency graph was constructed where nodes represent farms and edges connect farms that spatially touch. A breadth-first search algorithm was employed to group adjacent farms into "configurations". Each configuration aimed to contain a minimum of 2 and a maximum of 10 farms. To create a final, continuous landscape representation encompassing all configurations for visualization or area-wide analysis, a global gap-filling procedure was performed.

## Ecological Intensification (EI) Model

This section details the methodology for the farm-level EI optimization model, which constitutes the first stage of our proposed framework. The model aims to maximize the total farm Net Present Value (NPV) over a time horizon $T$ with a discount rate $r$, as formulated in the "Farm-Level EI Optimization Outcomes" section.

### Model Implementation

The optimization is implemented as a non-linear, continuous optimization problem using Pyomo. The IPOPT solver is employed to handle the non-linearities inherent in the yield enhancement functions. Essential spatial inputs, such as pairwise distances between plot centroids ($d_{ij}$) and neighbor sets ($N(i)$ for each plot $i$, defined as plots $j$ within a specified distance), are precomputed using the to improve efficiency.

### Decision Variables

For each plot $i$ identified as an agricultural plot ($i \in I_{ag}$), the model determines the optimal values for:

1. $m_i$: Continuous variable representing the fraction of plot $i$'s area allocated to field margin intervention ($0 \leq m_i \leq 1$).
2. $h_i$: Continuous variable representing the fraction of plot $i$'s area converted to habitat ($0 \leq h_i \leq 1$).

**NPV Calculation ($NPV_i$)**
The NPV for each plot $i$ is calculated based on the discounted cash flows over the time horizon $T$. The components differ slightly for agricultural plots versus existing habitat plots.

For Agricultural Plots ($i \in I_{ag}$), the NPV calculation involves initial implementation costs and annual net cash flows (revenue minus maintenance costs and yield loss from habitat conversion), discounted back to the present.

**Implementation Costs ($C_{impl,i}$)**
A one-time, undiscounted cost incurred at $t = 0$:

$$C_{impl,i} = A_i(m_i C_{m,impl} + h_i C_{h,impl})$$

where $A_i$ is the area of plot $i$, and $C_{m,impl}$ and $C_{h,impl}$ are the per-unit-area implementation costs for margin and habitat, respectively.

**Annual Maintenance Costs ($C_{maint,i}$)**
Costs incurred each year ($t = 1...T$):

$$C_{maint,i} = A_i(m_i C_{m,maint} + h_i C_{h,maint} + (1 - h_i) C_{ag,maint})$$

using per-unit-area maintenance costs for margin ($C_{m,maint}$), converted habitat ($C_{h,maint}$), and standard agriculture ($C_{ag,maint}$).

**Annual Revenue ($R_{i,t}$)**
Revenue generated from crop yield in year $t$:

$$R_{i,t} = Y_{i,t} \cdot p_c \cdot A_i \cdot (1 - h_i)$$

where $p_c$ is the crop price, $A_i$ is the plot area, $(1 - h_i)$ is the fraction of the plot remaining in production, and $Y_{i,t}$ is the combined yield per unit area in year $t$.

**Combined Yield ($Y_{i,t}$)**
The yield per unit area is enhanced by pollination and pest control services originating from interventions on plot $i$ and its neighbors $j \in N(i)$.

$$Y_{i,t} = Y_{base,i} \cdot (1 + P_{i,t} + S_{i,t})$$

where $Y_{base,i}$ is the baseline yield, $P_{i,t}$ is the fractional yield increase from pollination, and $S_{i,t}$ is the fractional yield increase from pest suppression at time $t$.

**Pollination ($P_{i,t}$) and Pest Suppression ($S_{i,t}$)**

These effects accumulate over time and decay with distance. They are calculated by summing the contributions from plot $i$ itself and its neighbors $j \in N(i)$ (including other agricultural plots and existing habitat plots. The effect of an intervention (margin $m_j$ or habitat $h_j$) on plot $j$ contributing to plot $i$ at time $t$ depends on the distance $d_{ij}$.

**Pollination Effect ($P_{i,t}$)**

$$P_{i,t} = \sum_{j \in N(i) \cup \{i\}} \left[ (\alpha m_j e^{-\beta d_{ij}})(1 - e^{-\gamma t}) + (\alpha_h h_j e^{-\beta_h d_{ij}})(1 - e^{-\gamma_h t}) \right] + \sum_{k \in N_{hab}(i)} (\alpha_h e^{-\beta_h d_{ik}})(1 - e^{-\gamma_h t})$$

**Pest Suppression Effect ($S_{i,t}$)**

$$S_{i,t} = \sum_{j \in N(i) \cup \{i\}} \left[ (\delta m_j e^{-\epsilon d_{ij}})(1 - e^{-\zeta t}) + (\delta_h h_j e^{-\epsilon_h d_{ij}})(1 - e^{-\zeta_h t}) \right] + \sum_{k \in N_{hab}(i)} (\delta_h e^{-\epsilon_h d_{ik}})(1 - e^{-\zeta_h t})$$

Here, $\alpha, \beta, \gamma$ and $\delta, \epsilon, \zeta$ are the strength, distance decay, and time accumulation parameters for pollination and pest control from margins, respectively. Parameters with subscript $h$ ($\alpha_h, \beta_h, ...\zeta_h$) represent the corresponding effects from habitat (converted or existing). $N_{hab}(i)$ is the set of neighboring plots that are existing habitats. All parameters are specific to the crop type on plot $i$. The time accumulation factors $(1 - e^{-(\cdot)t})$ are precomputed for efficiency.

**Total NPV for Agricultural Plot ($NPV_i$)**

$$NPV_i = -C_{impl,i} + \sum_{t=1}^{T} \frac{R_{i,t} - C_{maint,i}}{(1+r)^t}$$

**For Existing Habitat Plots ($k \in I_{hab}$)**

These plots incur a constant annual maintenance cost ($C_{exist,hab}$ and do not generate revenue, but they contribute to pollination and pest control services for neighboring agricultural plots.

$$NPV_k = \sum_{t=1}^{T} \frac{-A_k C_{exist,hab}}{(1+r)^t}$$

The model solves for the values of $m_i$ and $h_i$ for all $i \in I_{ag}$ that maximize the total farm NPV objective function These fractional allocations represent the economically optimal investment level in EI per plot, which are then passed to the next stage of the framework.

**Model Parameters**
The parameters used in the farm-level EI model are summarized in Table 1.

| Parameter Group | Parameter Definition | Symbol / Name | Value(s) | Unit / Notes |
|---|---|---|---|---|
| Crop-Specific | Spring wheat | | | |

| | Margin Pollination Strength | $\alpha$ | 0.05 | Unitless (yield fraction) |
|---|---|---|---|---|
| | Margin Pollination Distance Decay | $\beta$ | 0.01 | Per meter (m$^{-1}$) |
| | Margin Pollination Time Accumulation | $\gamma$ | 0.2 | Per year (yr$^{-1}$) |
| | Margin Pest Control Strength | $\delta$ | 0.05 | Unitless (yield fraction) |
| | Margin Pest Control Distance Decay | $\epsilon$ | 0.01 | Per meter (m$^{-1}$) |
| | Margin Pest Control Time Accumulation | $\zeta$ | 0.2 | Per year (yr$^{-1}$) |
| | Habitat Pollination Strength | $\alpha_h$ | 0.05 | Unitless (yield fraction) |
| | Habitat Pollination Distance Decay | $\beta_h$ | 0.005 | Per meter (m$^{-1}$) |
| | Habitat Pollination Time Accumulation | $\gamma_h$ | 0.2 | Per year (yr$^{-1}$) |
| | Habitat Pest Control Strength | $\delta_h$ | 0.05 | Unitless (yield fraction) |
| | Habitat Pest Control Distance Decay | $\epsilon_h$ | 0.005 | Per meter (m$^{-1}$) |
| | Habitat Pest Control Time Accumulation | $\zeta_h$ | 0.2 | Per year (yr$^{-1}$) |
| | Price | $p_c$ | 200 | USD / Tonne |
| **Barley** | | | | |
| | (parameters as above) | $\alpha \ldots \zeta_h$ | (0.05, 0.01, 0.2, 0.05, 0.01, 0.2, ... 0.2) | As above |
| | Price | $p_c$ | 120 | USD / Tonne |
| **Canola/rapeseed** | | | | |
| | (parameters as above) | $\alpha \ldots \zeta_h$ | (0.20, 0.01, 0.2, 0.10, 0.01, 0.2, 0.2) | As above |
| | Price | $p_c$ | 1100 | USD / Tonne |
| **Corn** | | | | |
| | (parameters as above) | $\alpha \ldots \zeta_h$ | (0.05, 0.01, 0.2, 0.05, 0.01, 0.2, 0.2) | As above |

|  | Price | $p_c$ | 190 | USD / Tonne |
|---|---|---|---|---|
|  | **Oats** |  |  |  |
|  | (parameters as above) | $\alpha \ldots \zeta_h$ | (0.05, 0.01, 0.2, 0.05, 0.01, 0.2, ... 0.2) | As above |
|  | Price | $p_c$ | 95 | USD / Tonne |
|  | **Soybeans** |  |  |  |
|  | (parameters as above) | $\alpha \ldots \zeta_h$ | (0.10, 0.01, 0.2, 0.10, 0.01, 0.2, 0.2) | As above |
|  | Price | $p_c$ | 370 | USD / Tonne |
| **General Economic** | Discount Rate | r | 0.05 | Unitless (5% per year) |
|  | Time Horizon | T | 20 | Years |
| **Costs** | **Margin Intervention** |  |  | Representative estimates |
|  | Implementation Cost | $C_{m,impl}$ | 400 | USD / ha (one-time) |
|  | Annual Maintenance Cost | $C_{m,maint}$ | 60 | USD / ha / year |
|  | **Habitat Conversion** |  |  |  |
|  | Implementation Cost | $C_{h,impl}$ | 300 | USD / ha (one-time) |
|  | Annual Maintenance Cost | $C_{h,maint}$ | 70 | USD / ha / year |
|  | **Existing Habitat** |  |  |  |
|  | Annual Maintenance Cost | $C_{exist,hab}$ | 0 | USD / ha / year |
|  | **Baseline Agriculture** |  |  |  |
|  | Annual Maintenance Cost | $C_{ag,maint}$ | 100 | USD / ha / year |

**Table 1: Parameters used in the farm-level Ecological Intensification (EI) model.** This table provides the comprehensive set of biophysical and economic parameters that drive the farm-level Ecological Intensification (EI) model. It is structured to perform a detailed cost-benefit analysis of implementing ecological features on agricultural land. For several key crops, the model quantifies the positive impact of ecosystem services, namely pollination and pest control, emanating from adjacent field margins and habitat patches. This effect on crop yield is simulated using crop-specific parameters for service strength, distance decay, and temporal accumulation, with certain crops like canola and soybeans modeled as more responsive. The entire analysis is framed within a standard economic structure, using a 20-year time horizon and a 5% annual discount rate to evaluate long-term profitability. Finally, the model incorporates a detailed cost structure, accounting for the one-time implementation and ongoing annual maintenance costs for both ecological interventions and baseline agriculture, weighed against the market prices of each crop.

The selection of parameter values is crucial for the model's outcomes.

*Crop-Specific Ecological Parameters ($\alpha, \beta, \gamma, \delta, \epsilon, \zeta$)*: These parameters quantify the strength, spatial reach, and temporal development of pollination and pest control services provided by margin and habitat interventions. Values were assigned based on available literature concerning insect dispersal distances, habitat effectiveness for beneficial insects, and crop-specific responses to these ecosystem services [7-10, 12-14]. Parameters for crops known to benefit significantly from insect pollination (Canola, Soybeans) have higher $\alpha$ values. Distance decay parameters ($\beta, \epsilon$) reflect the assumption that effects dissipate with distance from the intervention; the decay is assumed less rapid (smaller $\beta_h, \epsilon_h$) for larger habitat patches compared to linear margin features (larger $\beta, \epsilon$). Time accumulation parameters ($\gamma, \zeta$) represent the time required for the ecological functions to establish and are assumed constant across interventions in this parametrization. These values represent best estimates, acknowledging that precise field data can be highly variable.

*Crop Prices ($p_c$)*: Prices (USD/Tonne) were chosen to reflect representative market conditions prevailing around the study period. They aim to capture the relative profitability of different crops, which strongly influences economic optimisation. Market prices are inherently volatile, hence the importance of sensitivity analysis.

*General Economic Parameters ($r, T$)*: A discount rate ($r$) of 5% is standard in long-term agricultural and environmental economic assessments, representing the time value of money and opportunity cost of capital. A time horizon ($T$) of 20 years was selected as a common period for evaluating the returns on agricultural investments and the establishment of ecological benefits, balancing the capture of long-term effects against increasing uncertainty in distant projections.

*Cost Parameters ($C_{...}$)*: Implementation and maintenance costs (USD/ha) represent typical expenses associated with establishing and managing field margins, habitat patches, and standard agricultural practices. Costs directly influence the profitability of adopting EI measures. The maintenance cost for existing habitat ($C_{exist,hab}$) is set to 0, implying no additional yearly cost is assigned within the model for maintaining these areas, although they contribute ecological benefits. Baseline agricultural maintenance ($C_{ag,maint}$) represents standard operational costs. Recognizing the uncertainty and potential variability in these parameters, sensitivity analyses were conducted. This involved systematically varying key crop-specific ecological parameters, crop prices, and intervention costs to assess the robustness of the optimal solutions ($m_i, h_i$) and understand how farm-level economic outcomes depend on these assumptions.

## Ecological Connectivity (EC) Model

The Ecological Connectivity (EC) model is formulated as a mathematical optimization problem designed to identify the optimal placement of ecological interventions (field margins and habitat patches) to maximize landscape connectivity, subject to economic viability constraints for individual farms. The model is implemented using Pyomo.

**Objective Function**

Let $P$ be the set of all discrete ecological pieces, with $P_m \subset P$ denoting the subset of margin pieces, $P_h \subset P$ the subset of habitat cell pieces, and $P_{fh} \subset P$ the subset of existing full habitat plots. Let $A$ be the set of all pairs of pieces $(i, j)$ considered adjacent if their distance $d_{ij}$ is less than or equal to a predefined adjacency distance, $d_{adj}$.

The primary objective is to maximize a landscape connectivity metric, $Z$:

$$\text{Maximize } Z = \sum_{i \in P} s_i x_i + \sum_{(i,j) \in A} w_{ij} y_{ij}$$

Where:

- $x_i$ is a binary decision variable: $x_i = 1$ if piece $i$ is selected, 0 otherwise. For pieces in $P_{fh}$, $x_i$ is fixed to 1.

- $s_i$ is the intrinsic ecological score of piece $i$. This score is defined based on the type of piece:

    - For habitat cells and full habitat plots ($i \in P_h \cup P_{fh}$): $s_i = a_i$, where $a_i$ is the area of piece $i$.

    - For margin arcs ($i \in P_m$): $s_i = w_m l_i$, where $l_i$ is the length of piece $i$, and $w_m$ is a weighting factor for margins.

- $y_{ij}$ is a binary decision variable: $y_{ij} = 1$ if both piece $i$ and piece $j$ are selected and are adjacent, 0 otherwise.

- $w_{ij}$ is the weight of the connection between adjacent selected pieces $i$ and $j$. For an IIC-like metric, this can be a function of their respective areas ($a_i, a_j$) and lengths ($l_i, l_j$), potentially including an area-length interaction factor ($\alpha_{AL}$):

$$w_{ij} = (l_i l_j) + (a_i a_j) + \alpha_{AL}(a_i l_j + a_j l_i)$$

The Pyomo implementation constructs this objective function by summing terms corresponding to individual piece scores and weighted adjacencies.

**Inputs**

Key inputs to this model include: a) Geospatial data for farm and plot boundaries. b) Economic parameters ($\Pi$): including costs for implementing and maintaining interventions, crop prices, discount rates, and parameters defining yield responses to ecological services. c) Baseline Net Present Value for each farm ($N_{f,base}$), where $f$ is the farm index. This baseline is calculated using initial intervention levels derived from the preceding EI model. d) Maximum allowable NPV loss ratio ($\rho_{max}$). e) Discretization parameters: number of boundary segments per plot for margins ($S_b$) and target number of interior cells per plot for habitat ($S_c$). e) Connectivity parameters: adjacency distance ($d_{adj}$) and the area-length interaction factor ($\alpha_{AL}$). f) Neighborhood distance ($d_{neib}$) for yield effects.

**Discretization of Landscape**

The landscape is discretized into candidate pieces: a) Margins: Boundaries of agricultural plots are divided into $S_b$ linear segments (arcs). b) Habitat Patches: Interiors of agricultural plots are partitioned into approximately $S_c$ polygonal cells using Voronoi tessellation. c) Full Habitat: Existing habitat plots ($P_{fh}$) are treated as single, indivisible pieces. d) Agricultural Interior Reference: A point (e.g., centroid) represents each agricultural plot's interior for calculating localized yield effects.

**Decision Variables**

The primary decision variables are: a) $x_i$: Binary, 1 if candidate margin arc or habitat cell $i$ is selected, 0 otherwise. As noted, $x_i = 1$ for $i \in P_{fh}$. b) $y_{ij}$: Binary, 1 if a connection between selected adjacent pieces $i$ and $j$ is formed, 0 otherwise.

**Constraints**

*Intervention Fraction Definition*: The selection of pieces implicitly defines the effective intervention fractions for each plot $p$. Let $M_p \subseteq P_m$ be the set of margin pieces on plot $p$, and $H_p \subseteq P_h$ be the set of habitat pieces on plot $p$. Let $P_{geom,p}$ be the perimeter of plot $p$ and $A_{geom,p}$ be its area. The effective margin fraction for plot $p$, $F_{m,p}$, is:

$$F_{m,p} = \frac{\sum_{i \in M_p} l_i x_i}{P_{geom,p}}$$

The effective habitat fraction for plot $p$, $F_{h,p}$, is:

$$F_{h,p} = \frac{\sum_{i \in H_p} a_i x_i}{A_{geom,p}}$$

These fractions influence the economic calculations.

*Economic Constraint (NPV Maintenance)*: For each farm $f$, its recalculated Net Present Value ($N_{f,new}$) must satisfy:

$$N_{f,new} \geq (1 - \rho_{max}) N_{f,base}$$

The term $N_{f,new}$ is the sum of individual plot NPVs ($N_{p,new}$) within farm $f$. Each $N_{p,new}$ is a function of implementation costs, maintenance costs, revenue from yield, and yield loss from habitat conversion. Revenue is affected by a plot-specific yield factor ($YF_p$), which incorporates pollination and pest control services from selected margin and habitat pieces within a neighborhood distance ($d_{neib}$). These ecological service effects are modeled using exponential decay functions based on distance to the service-providing pieces. The inclusion of these terms makes the NPV calculation non-linear.

*Adjacency Constraints*: For pairs of pieces $(i, j)$ (excluding agricultural interior reference points) that are within $d_{adj}$ of each other: If piece $i \notin P_{fh}$ and piece $j \notin P_{fh}$:

$$y_{ij} \leq x_i \quad y_{ij} \leq x_j \quad y_{ij} \geq x_i + x_j - 1 \quad \text{If piece } i \in P_{fh} \text{ (so } x_i = 1\text{) and piece } j \notin P_{fh}:$$

$$y_{ij} = x_j \quad \text{If both } i, j \in P_{fh}:$$

$y_{ij} = 1$ (if they are considered adjacent based on $d_{adj}$)

**Implementation and Solver**
The model is constructed as a ConcreteModel in Pyomo. A distance matrix $D$, containing all pairwise distances $d_{ij}$ between piece geometries, is pre-calculated. Given the non-linearities in the NPV constraint expressions (arising from exponential functions and products of variables in yield factor calculations), a solver suitable for Mixed-Integer Non-Linear Programming (MINLP) is employed, such as IPOPT.

The model's primary output is the set of selected margin arcs and habitat cells (where $x_i = 1$), defining the optimized spatial configuration of ecological interventions. It also provides the maximized connectivity value ($Z$) and the recalculated NPVs for plots and farms.

**Model Hyperparameters**

The hyperparameters used in the landscape-level EC model are summarized in Table 2.

| Parameter Symbol | Description | Value | Justification |
|---|---|---|---|
| $S_b$ | Boundary segments per plot | 4 | Provides a basic level of spatial choice for margin placement (e.g., on different sides of a plot) while keeping the number of decision variables manageable. |
| $S_c$ | Interior cells per plot | 4 | Offers a simplified representation of potential habitat patches within plots (e.g., in different quadrants), balancing some spatial flexibility with computational tractability. |
| $d_{adj}$ | Adjacency distance | 0.0 m | Defines the strictest condition for direct structural connectivity, requiring pieces to physically touch. |
| Metric | Connectivity metric | IIC | The Integral Index of Connectivity (IIC) is chosen as it's a robust and widely accepted metric in landscape ecology for evaluating habitat availability and reachability. It accounts for both patch areas and the connections between them, making it sensitive to habitat fragmentation. |
| $\alpha_{AL}$ | Area-length interaction factor (for IIC) | $10^{-9}$ | Minimizes the influence of these mixed-attribute terms. Direct connections are primarily valued based on interactions between similar feature types (area-area for habitats, length-length for margins). It also serves to prevent margins from lining along habitats, and instead extend them. |
| $\rho_{max}$ | Maximum NPV loss ratio | 0.2 | Represents an acceptable economic trade-off, allowing up to a 20% reduction in a farm's baseline Net Present Value to achieve improved landscape connectivity. This value balances ecological ambition with economic feasibility for farmer participation and is typically informed by policy or stakeholder considerations. |
| $d_{neib}$ | Neighborhood distance for yield effects | 1000 m | Defines the spatial extent (1 km) over which ecosystem services (e.g., pollination, pest control) from an intervention are assumed influence crop yields. This distance is a generalization based on the typical foraging or dispersal ranges of relevant beneficial organisms in agricultural settings. |
| $\epsilon_{sol}$ | Solver exit tolerance (optimality gap) | $10^{-6}$ | This is a standard precision level for numerical optimization solvers (e.g., IPOPT for non-linear problems or CBC for linear problems). It defines the stopping criterion related to the quality optimality of the solution found (such as the satisfaction of Karush-Kuhn-Tucker conditions for non-linear problems, or the relative gap for mixed-integer linear programs). The value offers a good compromise between solution accuracy and the computational time required, especially for complex models. |
| $w_m$ | Margin weight in objective's score term | 50 | This weighting factor scales the contribution of linear margin elements relative to areal habitat elements when calculating the intrinsic ecological score of individual pieces in the objective function. Such a high weight is justified as margins are exceptionally effective or critical for landscape connectivity per unit length (e.g., serving as vital corridors for specific target species). |
| $\Pi$ | Economic & Biophysical Parameters | External | This represents a comprehensive set of site-specific parameters (e.g., costs, prices, crop yields, coefficients for yield impact functions like $\alpha, \beta, \gamma, \delta, \epsilon, \zeta$) derived from literature. These are typically calibrated from local agricultural data, economic survey and ecological literature. |

**Table 2: Hyperparameters used in the landscape-level EC model.** This table specifies the key parameters that define the structure, objectives, and constraints of the landscape-level optimization model. The model's spatial framework simplifies agricultural plots into a manageable number of discrete boundary and interior segments. The primary ecological objective is to maximize habitat connectivity, which is quantified using the Integral Index of Connectivity (IIC) under a strict "touching" adjacency rule. This ecological calculation is tuned, with specific weights that prioritize the contribution of linear margins to the overall connectivity score. The search for optimal landscape designs is conducted within realistic constraints, including a cap on acceptable economic impact, limiting any solution to a maximum of 20% Net Present Value (NPV) loss for the farm. Furthermore, the model incorporates biophysical assumptions, such as a 1 km spatial scale for the effect of ecosystem services on crop yields. The framework is parameterized by a comprehensive set of external economic and ecological data derived from relevant literature.

It is important to recognize that the resulting landscape connectivity configurations are sensitive to the specific values chosen for parameters such as the Area-Length Interaction Factor ($\alpha_{AL}$) and and the Margin Weight ($w_m$). The EC model is designed with the flexibility to accommodate varying values for these and other parameters, allowing for calibration and sensitivity analyses. For practical application, these parameters should be carefully considered and potentially tuned to align with local ecological contexts, specific conservation objectives for target species or processes, and regional land management priorities.

## Bayesian Optimization (BO) to Find Policy Instruments

The goal of the BO process is to identify a set of parameters $\theta_P$ for farm-level policy instruments $P$ such that farmers, responding to these policies, make land-use decisions that result in landscape connectivity ($C_{L,k}^{policy}$) and farm NPVs ($NPV_{f,k}^{policy}$) that are as close as possible to the target values ($C_{L,k}^{target}$, $NPV_{f,k}^{target}$) across various landscape configurations.

**Policy Instruments ($P(\theta_P)$)**

The vector of policy parameters $\theta_P$ can include:

1. $s_{est,m}, s_{est,h}$: Subsidy rates for the establishment of margins and habitats adjacent to existing habitats.
2. $s_{maint,m}, s_{maint,h}$: Subsidy rates for the maintenance of margins and habitats.
3. $p_{ha}$: Direct payment per hectare for land converted to habitat.
4. $m_{area}$: Mandated minimum total area of habitat per farm.
5. $m_{frac}$: Mandated minimum fraction of a field margin to be intervened.
6. $e_c$: Eco-premium factor for crop $c$, applied as a multiplier to the base price of crops.

**Bayesian Optimization Process**

The Bayesian Optimization (BO) algorithm iteratively selects candidate sets of policy parameters, denoted as $\theta_P$, from the defined multi-dimensional search space for evaluation. The core of the BO process lies in quantifying the performance of each selected $\theta_P$. This evaluation is designed to assess how effectively a policy $P(\theta_P)$ achieves the desired landscape and farm outcomes, and involves the following steps for each candidate $\theta_P$. First, to ensure the robustness and generalizability of the policy, its performance is assessed across a sample of $N_S$ distinct landscape configurations. For each landscape configuration $k$ (where $k = 1, \ldots, N_S$) within this sample, the following sequence of analyses is performed:

*Farm-Level Economic Response under Policy*: The introduction of policy $P(\theta_P)$ alters the economic optimization problem faced by each farm $f$ within the configuration $k$. The original farm-level objective (e.g., maximizing Net Present Value, $NPV_f$) is modified to incorporate the financial implications of the policy. This involves adjusting revenues based on eco-premiums (e.g., crop revenue $R_c$ becomes $R_c \cdot e_c$), reducing costs due to subsidies (e.g., habitat maintenance cost $C_{maint}$ becomes $C_{maint} \cdot (1 - s_{maint})$), accounting for direct payments (e.g., $p_{ha}$ for habitat area), and ensuring compliance with any mandated ecological measures (e.g., $m_{area}, m_{frac}$). The farm $f$ then re-optimizes its land allocations to these new economic conditions, resulting in a set of policy-induced land use decisions, $x_{f,k}^{policy}(\theta_P)$, and a new farm-level Net Present Value, $NPV_{f,k}^{policy}(\theta_P)$. The direct financial expenditure incurred by the implementing agency for farm $f$ due to policy $P(\theta_P)$ in configuration $k$ is also quantified as $PC_{f,k}(\theta_P)$.

*Aggregation and Landscape-Level Connectivity Assessment*: The ecological intervention areas resulting from the policy-induced land allocations $x_{f,k}^{policy}(\theta_P)$ from all farms $f \in F_k$ (the set of farms in configuration $k$) are aggregated for the entire landscape configuration. These aggregated habitat areas are then subject to a spatial optimization procedure (an ecological connectivity repositioning model). This procedure determines the optimal spatial arrangement of these policy-induced habitat amounts to maximize landscape connectivity, yielding the policy-induced landscape connectivity score, $C_{L,k}^{policy}(\theta_P)$, for configuration $k$. The Integral Index of Connectivity ($IIC$) is the metric used for $C_{L,k}^{policy}$.

*Configuration-Specific Policy Cost*: The total financial expenditure for implementing policy $P(\theta_P)$ across all farms in configuration $k$ is calculated as $PC_k(\theta_P) = \sum_{f \in F_k} PC_{f,k}(\theta_P)$.

Once these values, $C_{L,k}^{policy}(\theta_P)$, $NPV_{f,k}^{policy}(\theta_P)$ for all $f \in F_k$, and $PC_k(\theta_P)$, have been determined for all $N_S$ sampled landscape configurations, they are used to calculate the overall objective function value, $O_{BO}(\theta_P)$, for the candidate policy parameter set $\theta_P$, as defined in the subsequent "Objective Function for BO" subsection. This objective value $O_{BO}(\theta_P)$ then informs the BO algorithm's strategy for selecting subsequent candidate parameter sets $\theta_P$ to evaluate in its search for the optimal policy.

**Objective Function for BO ($O_{BO}$)**
The function to be minimized by the BO algorithm is the average weighted sum of deviations from target connectivity and NPV, across the sampled configurations:

$$O_{BO}(\theta_P) = \frac{1}{N_S} \sum_{k=1}^{N_S} \left( w_C \cdot |C_{L,k}^{policy}(\theta_P) - C_{L,k}^{target}| + w_{NPV} \cdot \left( \frac{1}{|F_k|} \sum_{f \in F_k} |NPV_{f,k}^{policy}(\theta_P) - NPV_{f,k}^{target}| \right) \right)$$

where $w_C$ and $w_{NPV}$ are weights, and $F_k$ is the set of farms in configuration $k$.

A penalty is applied if the average policy cost across sampled configurations exceeds a predefined maximum budget ($B_{max}$). If $\left( \frac{1}{N_S} \sum_{k=1}^{N_S} PC_k(\theta_P) \right) > B_{max}$, then $O_{BO}(\theta_P)$ is assigned a large penalty value (e.g., $10^{15}$).

This detailed methodology allows for the systematic identification of effective, non-spatially explicit policy instruments that can guide agricultural landscapes towards enhanced ecological connectivity while considering economic realities at the farm level.

The BO explores a defined hyperrectangle for $\theta_P$, where each policy parameter has a specified range which is detailed in Table 3.

| Policy Parameter Description | Symbol | Search Range (Min, Max) | Units |
|---|---|---|---|
| Subsidy factor for margin establishment adjacent to existing habitats | $s_{est,m}$ | [0.0, 0.5] | Dimensionless |
| Subsidy factor for habitat establishment adjacent to existing habitats | $s_{est,h}$ | [0.0, 0.5] | Dimensionless |
| Subsidy factor for margin maintenance | $s_{maint,m}$ | [0.0, 0.5] | Dimensionless |
| Subsidy factor for habitat maintenance | $s_{maint,h}$ | [0.0, 0.5] | Dimensionless |
| Payment per hectare for habitat conversion | $p_{ha}$ | [0, 150] | Currency units/ha |
| Mandated minimum total habitat area per farm | $m_{area}$ | [0, 10] | ha |

| Mandated minimum fraction of margin adjacent to existing habitats | $m_{frac}$ | [0.0, 0.3] | Dimensionless |
|---|---|---|---|
| Eco-premium factor for crop c | $e_c$ | [1.0, 1.3] | Dimensionless |

**Table 3: Ranges of the Policy Parameter Search Space.** This table details the suite of policy instruments investigated in this study, defining the search space for the optimization algorithm. These eight parameters represent a diverse set of governmental interventions aimed at influencing land-use decisions to achieve conservation objectives and can be broadly categorized into three types: financial incentives (subsidy factors $s_{est,m}$, $s_{est,h}$, $s_{maint,m}$, $s_{maint,h}$ and a direct payment $p_{ha}$), regulatory mandates (a minimum habitat area $m_{area}$ and an adjacency fraction $m_{frac}$), and market-based mechanisms (an eco-premium factor $e_c$). The search range for each parameter defines the lower and upper bounds explored by the Bayesian Optimization algorithm, representing plausible policy scenarios from no intervention to significant pressure. The optimization process searches within this multi-dimensional space to identify the combination of policy settings that yields the most desirable outcomes.

## Model Hyperparameters

The hyperparameters used in the BO are summarized in Table 4. These hyperparameters primarily influence the search strategy, computational budget, and objective definition

| Hyperparameter | Symbol | Value | Justification |
|---|---|---|---|
| Connectivity Weight | $w_C$ | 1 | Assigns weight to the connectivity difference term. |
| NPV Weight | $w_{NPV}$ | 0 | Assigns weight to the farm NPV difference term in the objective. A value of 0 indicates that deviations in farm NPV are not directly penalized during the BO's search for optimal policy parameters. |
| Configuration Samples | $N_S$ | 20 | The number of distinct landscape configurations sampled to evaluate each candidate policy parameter set during one BO iteration. The chosen value aims to provide a robust estimate of a policy's average performance across diverse contexts, balancing evaluation fidelity with computational load. |
| BO Initial Points | $N_{init}$ | 15 | The number of candidate policy parameter sets evaluated with a space-filling (e.g., random or Latin hypercube) sampling strategy before the Gaussian Process model actively guides the BO search. The chosen value provides an initial exploration of the search space to build a preliminary model of the objective function. |
| BO Total Calls (Budget) | $N_{calls}$ | 100 | The total number of evaluations of the objective function allowed for the entire BO process, inclusive of the initial points. It defines the computational budget for the optimization, representing a trade-off between search thoroughness and practical time constraints. |

**Table 4: Bayesian Optimization (BO) Hyperparameter Configuration.** This table outlines the specific hyperparameter settings used to guide the Bayesian Optimization process. The objective function is configured to exclusively prioritize the connectivity goal, with the connectivity weight ($w_C$) set to 1 and the Net Present Value (NPV) weight ($w_{NPV}$) set to 0. This means the optimization algorithm actively searches for policy parameters that improve landscape connectivity, while farm NPV is treated as an observed outcome rather than a direct optimization target. To ensure the evaluation of each policy is robust and not biased by a single scenario, its performance is averaged across $N_S$=20 distinct landscape configuration samples. The BO search itself is constrained by a total computational budget of $N_{calls}$=100 iterations. The search begins with an initial exploration phase of $N_{init}$=15 points selected via a space-filling sampling method to build a preliminary surrogate model of the objective function. The subsequent 85 iterations are then actively guided by the BO algorithm to efficiently locate the optimal policy parameter set within the defined budget.

These hyperparameter settings are crucial for tailoring the BO search to the specific problem of finding effective and economically mindful ecological policies. The choice of $w_C = 1$ and $w_{NPV} = 0$ specifically directs the optimization towards achieving ecological connectivity targets as the primary objective, within the defined policy budget $B_{max}$ (as specified in the $O_{BO}(\theta_P)$ definition). The sampling parameters ($N_S, N_{init}, N_{calls}$) are chosen to balance the depth of exploration and exploitation of the policy parameter space with the significant computational cost associated with each evaluation of $O_{BO}(\theta_P)$.

### Synthetic Data Experiments

Synthetic farm and plot-level geospatial data were generated to simulate agricultural landscapes. A total of 500 configurations were generated. For each configuration, the process involved defining overall farm-cluster boundaries and subsequently partitioning this area into a specific number of farms using Voronoi tessellation. A Voronoi diagram was constructed based on random points generated within the boundary, and the resulting Voronoi cells were clipped to the boundary. This procedure generated a set of non-overlapping farms that collectively covered the chosen area. The number of farms ranged from five to ten. Each individual farm polygon was then subjected to a partitioning process to create internal subplots. This subdivision was also achieved using Voronoi tessellation. This procedure generated a set of smaller, non-overlapping polygons (plots) that collectively covered the original farm polygon. The number of plots ranged from five to ten.

Following the geometric partitioning, attributes were assigned to each plot. Plots were first randomly categorized into primary types, such as 'agricultural plot' or 'habitat plot', based on predefined probabilities. Sampling probabilities of 0.6 and 0.4 were assigned to them respectively in order to have slightly more agricultural plots than habitat plots. Subsequently, a specific label was assigned to each plot, again through a weighted random selection process dependent on its primary type. Agricultural plots received labels corresponding to crop types ("Soybeans", "Oats", "Corn", "Canola/rapeseed", "Barley", "Spring wheat"), while habitat plots received labels reflecting land cover types ("Broadleaf", "Coniferous", "Exposed land/barren", "Grassland", "Shrubland", "Water", "Wetland"). The weights used for this random assignment were intended to reflect typical distributions of these categories in the Canadian annual crop inventory (CACI) of 2022 [47]. Additionally, a numerical yield value was randomly assigned to each plot, drawn from a distribution as seen in the CACI for the crop in the plot. The final output for each synthetic farm consisted of a GeoJSON FeatureCollection containing the geometric definitions (polygons) and assigned properties (type, label, yield) for all its constituent plots. This process resulted in, for each of the 500 configurations, a structured dataset of synthetic farms composed of multiple labelled plots mimicking real-world agricultural field heterogeneity. See Synthetic Data section in supplementary and supplementary Fig. 14-17.

# References

1. Batáry, P., Dicks, L. V., Kleijn, D., & Sutherland, W. J. (2015). The role of agri‑environment schemes in conservation and environmental management. *Conservation Biology*, *29*(4), 1006-1016.
2. Scherr, S. J., & McNeely, J. A. (2008). Biodiversity conservation and agricultural sustainability: towards a new paradigm of 'ecoagriculture'landscapes. *Philosophical Transactions of the Royal Society B: Biological Sciences*, *363*(1491), 477-494.
3. Tittonell, P. (2014). Ecological intensification of agriculture—sustainable by nature. *Current Opinion in Environmental Sustainability*, *8*, 53-61.
4. Boetzl, F. A., Krauss, J., Heinze, J., Hoffmann, H., Juffa, J., König, S., ... & Steffan-Dewenter, I. (2021). A multitaxa assessment of the effectiveness of agri-environmental schemes for biodiversity management. *Proceedings of the National Academy of Sciences*, *118*(10), e2016038118.
5. Kremen, C. (2020). Ecological intensification and diversification approaches to maintain biodiversity, ecosystem services and food production in a changing world. *Emerging Topics in Life Sciences*, *4*(2), 229-240.
6. Petersen, B., & Snapp, S. (2015). What is sustainable intensification? Views from experts. *Land use policy*, *46*, 1-10.
7. Pywell, R. F., Heard, M. S., Woodcock, B. A., Hinsley, S., Ridding, L., Nowakowski, M., & Bullock, J. M. (2015). Wildlife-friendly farming increases crop yield: evidence for ecological intensification. *Proceedings of the Royal Society B: Biological Sciences*, *282*(1816), 20151740.
8. Garibaldi, L. A., Steffan-Dewenter, I., Winfree, R., Aizen, M. A., Bommarco, R., Cunningham, S. A., ... & Klein, A. M. (2013). Wild pollinators enhance fruit set of crops regardless of honey bee abundance. *science*, *339*(6127), 1608-1611.
9. Morandin, L. A., & Kremen, C. (2013). Hedgerow restoration promotes pollinator populations and exports native bees to adjacent fields. *Ecological Applications*, *23*(4), 829-839.
10. Hatt, S., Boeraeve, F., Artru, S., Dufrêne, M., & Francis, F. (2018). Spatial diversification of agroecosystems to enhance biological control and other regulating services: An agroecological perspective. *Science of the Total Environment*, *621*, 600-611.
11. Kleijn, D., Bommarco, R., Fijen, T. P., Garibaldi, L. A., Potts, S. G., & Van Der Putten, W. H. (2019). Ecological intensification: bridging the gap between science and practice. *Trends in ecology & evolution*, *34*(2), 154-166.
12. Dainese, M., Montecchiari, S., Sitzia, T., Sigura, M., & Marini, L. (2017). High cover of hedgerows in the landscape supports multiple ecosystem services in Mediterranean cereal fields. *Journal of Applied Ecology*, *54*(2), 380-388.
13. Albrecht, M., Kleijn, D., Williams, N. M., Tschumi, M., Blaauw, B. R., Bommarco, R., ... & Sutter, L. (2020). The effectiveness of flower strips and hedgerows on pest control, pollination services and crop yield: a quantitative synthesis. *Ecology letters*, *23*(10), 1488-1498.
14. Woodcock, B. A., Bullock, J. M., McCracken, M., Chapman, R. E., Ball, S. L., Edwards, M. E., ... & Pywell, R. F. (2016). Spill-over of pest control and pollination services into arable crops. *Agriculture, Ecosystems & Environment*, *231*, 15-23.
15. Merckx, T., & Pereira, H. M. (2015). Reshaping agri-environmental subsidies: From marginal farming to large-scale rewilding. *Basic and Applied Ecology*, *16*(2), 95-103.

16. López-Cubillos, S., McDonald-Madden, E., Mayfield, M. M., & Runting, R. K. (2023). Optimal restoration for pollination services increases forest cover while doubling agricultural profits. *PLoS Biology*, *21*(5), e3002107.
17. López-Cubillos, S., Runting, R. K., Mayfield, M. M., & Mcdonald-Madden, E. (2021). Catalytic potential of pollination services to reconcile conservation and agricultural production: a spatial optimization framework. *Environmental Research Letters*, *16*(6), 064098.
18. Magrach, A., Champetier, A., Krishnan, S., Boreux, V., & Ghazoul, J. (2019). Uncertainties in the value and opportunity costs of pollination services. *Journal of Applied Ecology*, *56*(7), 1549-1559.
19. Montoya, D., Haegeman, B., Gaba, S., De Mazancourt, C., Bretagnolle, V., & Loreau, M. (2019). Trade‑offs in the provisioning and stability of ecosystem services in agroecosystems. *Ecological Applications*, *29*(2), e01853.
20. Guo, S., Saito, K., Yin, W., & Su, C. (2018). Landscape connectivity as a tool in green space evaluation and optimization of the Haidan District, Beijing. *Sustainability*, *10*(6), 1979.
21. Kukkala, A. S., & Moilanen, A. (2017). Ecosystem services and connectivity in spatial conservation prioritization. *Landscape Ecology*, *32*, 5-14.
22. Acevedo-Osorio, Á., Cárdenas, J. S., & Martín-Pérez, A. M. (2024). Agroecological planning of productive systems with functional connectivity to the ecological landscape matrix: two Colombian case studies. *Frontiers in Sustainable Food Systems*, *8*, 1257540.
23. Bird, K. I., Uden, D. R., & Allen, C. R. (2023). Functional connectivity varies across scales in a fragmented landscape. *PloS one*, *18*(8), e0289706.
24. Pither, R., O'Brien, P., Brennan, A., Hirsh-Pearson, K., & Bowman, J. (2023). Predicting areas important for ecological connectivity throughout Canada. *PLoS One*, *18*(2), e0281980.
25. Catchen, M. D., Lin, M., Poisot, T., Rolnick, D., & Gonzalez, A. (2023). Improving ecological connectivity assessments with transfer learning and function approximation.
26. Koh, I., Rowe, H. I., & Holland, J. D. (2013). Graph and circuit theory connectivity models of conservation biological control agents. *Ecological Applications*, *23*(7), 1554-1573.
27. Pinto, N., & Keitt, T. H. (2009). Beyond the least-cost path: evaluating corridor redundancy using a graph-theoretic approach. *Landscape Ecology*, *24*, 253-266.
28. Urban, D., & Keitt, T. (2001). Landscape connectivity: a graph‑theoretic perspective. *Ecology*, *82*(5), 1205-1218.
29. Ewer, T., Smith, T., Cook, S., Jones, S., DeClerck, F., & Ding, H. (2023). Aligning regenerative agricultural practices with outcomes to deliver for people, nature and climate. *The Food and Land Use Coalition.* https://www. foodandlandusecoalition. org/wp-content/uploads/2023/01/Aligning-regenerative-agricultural-practiceswith-outcomes-to-deliver-for-people-nature-climate-Jan-2023. pdf.
30. Petit, S., & Landis, D. A. (2023). Landscape-scale management for biodiversity and ecosystem services. *Agriculture, Ecosystems & Environment*, *347*, 108370.
31. Verhagen, W., van der Zanden, E. H., Strauch, M., van Teeffelen, A. J., & Verburg, P. H. (2018). Optimizing the allocation of agri-environment measures to navigate the trade-offs between ecosystem services, biodiversity and agricultural production. *Environmental Science & Policy*, *84*, 186-196.
32. Ververidis, K. A. (2008). A multi-objective bi-level optimisation model for agricultural policy in Scotland.

33. Weerasena, L., Shier, D., Tonkyn, D., McFeaters, M., & Collins, C. (2023). A sequential approach to reserve design with compactness and contiguity considerations. *Ecological Modelling*, *478*, 110281.
34. Chopin, P., Blazy, J. M., & Doré, T. (2015). A new method to assess farming system evolution at the landscape scale. *Agronomy for sustainable development*, *35*, 325-337.
35. Kuhfuss, L., Begg, G., Flanigan, S., Hawes, C., & Piras, S. (2019). Should agri-environmental schemes aim at coordinating farmers' pro-environmental practices? A review of the literature.
36. Zindler, M., Haensel, M., Fricke, U., Schmitt, T. M., Tobisch, C., & Koellner, T. (2024). Improving agri-environmental schemes: suggestions from farmers and nature managers in a Central European region. *Environmental Management*, *73*(4), 826-840.
37. Paul, C., Reith, E., Salecker, J., & Knoke, T. (2019). How integrated ecological-economic modelling can inform landscape pattern in forest agroecosystems. *Current Landscape Ecology Reports*, *4*, 125-138.
38. Brady, M., Sahrbacher, C., Kellermann, K., & Happe, K. (2012). An agent-based approach to modeling impacts of agricultural policy on land use, biodiversity and ecosystem services. *Landscape Ecology*, *27*, 1363-1381.
39. Geissler, C. H., Haan, N. L., Basso, B., Fowler, A., Landis, D. A., Lark, T. J., & Maravelias, C. T. (2025). A multi-objective optimization model for cropland design considering profit, biodiversity, and ecosystem services. *Ecological Modelling*, *500*, 110954.
40. Wätzold, F., & Drechsler, M. (2005). Spatially uniform versus spatially heterogeneous compensation payments for biodiversity-enhancing land-use measures. *Environmental and Resource Economics*, *31*, 73-93.
41. Armsworth, P. R., Daily, G. C., Kareiva, P., & Sanchirico, J. N. (2006). Land market feedbacks can undermine biodiversity conservation. *Proceedings of the National Academy of Sciences*, *103*(14), 5403-5408.
42. Hodge, I. (2000). Agri-environmental policy: a UK perspective. *Environmental Policy: Objectives, instruments and implementation*, 216-239.
43. Shahriari, B., Swersky, K., Wang, Z., Adams, R. P., & De Freitas, N. (2015). Taking the human out of the loop: A review of Bayesian optimization. *Proceedings of the IEEE*, *104*(1), 148-175.
44. Liczner, A. R., Pither, R., Bennett, J. R., Bowman, J., Hall, K. R., Fletcher Jr, R. J., ... & Pither, J. (2024). Advances and challenges in ecological connectivity science. *Ecology and Evolution*, *14*(9), e70231.
45. Will, M., Bartkowski, B., Schwarz, N., Wittstock, F., Grujic, N., Li, C., ... & Müller, B. (2024). From primary data to formalized decision-making: open challenges and ways forward to inform representations of farmers' behavior in agent-based models. *Ecology and society*, *29*(4), 31.
46. Ravaioli, G., Domingos, T., & Teixeira, R. F. (2023). A framework for data-driven agent-based modelling of agricultural land use. *Land*, *12*(4), 756.
47. Annual Crop Inventory 2022. Available Online: https://open.canada.ca/data/en/dataset/ee44d2d4-f887-47d1-9a5b-4c9c7b43bb46
48. Biomass Inventory Mapping and Analysis. Available Online: https://open.canada.ca/data/en/dataset/1a759d95-3008-4078-87af-5bb1bdf657b3

49. Dsouza, K. B. (2025). Connexus: Bridging farm economics and landscape ecology for global sustainability through hierarchical and Bayesian optimization [Source code]. GitHub. https://github.com/kevinbdsouza/Connexus.

## Data Availability

The data that support the findings of this study are publicly available to download and are referenced in the bibliography. Refer to the Methods section for more details. All project specific data generating scripts are made available in the GitHub repository [49].

## Code Availability

The code repository for this project, including data processing, optimization models, simulation workflows, and visualization scripts are available in the GitHub repository [49].

## Acknowledgements

We thank Mitacs and Royal Bank of Canada for primarily funding this research through the Accelerate program (K.B.D.). In addition, we acknowledge support from the Natural Sciences and Engineering Research Council of Canada (NSERC) Alliance Mission grant - ALLRP 577126-2022 (Y.L. and J.M.-C.), Canada Research Chairs Program - CRC-2023-00181 (J.M-.C.), and NSERC PDF program (K.B.D.).

## Authors' contributions

K.B.D. conducted the research, performed simulations, wrote the article, plotted results, and created illustrations. G.A.W., J.M.-C., and Y.L. were involved in the acquisition of funding, editing the article and overall supervision. All authors contributed to discussion and conceptualization of arguments.

## Competing interests

The authors declare no competing interests

# Supplementary

While the EI model optimizes purely for the NPV summed across all farm plots, the underlying economic logic creates emergent, non-random spatial patterns in the recommended interventions. The bar chart summarizing Pearson correlation coefficients (Fig. 1) quantifies the strength and direction of relationships between key plot factors and the optimal levels of margin and habitat interventions. The analysis revealed weak relationships between plot characteristics and the optimal intervention strategies selected by the model. Notably, plot yield exhibited a slight negative correlation (r=−0.12) with the decision to implement margins (Fig. 1), however, this correlation seems superficial according to the scatter plot (Fig. 2a). Geometric factors display weak negative correlations with habitat conversion (r ≈ -0.11 to -0.15). This suggests that larger or more geometrically complex plots are not favoured for habitat conversion in this optimization (Fig. 1, Fig. 2b). While costs scale with area in the model, the benefits generated are influenced by the intervention fraction and proximity to neighbours. The negative correlation might arise if the increasing costs on larger plots slightly outweigh the benefits, or if smaller plots are strategically located such that their conversion offers higher marginal NPV gains to the overall farm system. Moreover, the hypothesis that habitat interventions are directed to where marginal gains are highest (further from existing habitats) is not strongly supported by these outcomes (Fig. 1). This could be a result of multiple habitats existing in the landscape in close proximity, thereby diluting the effect of distance to existing habitat.

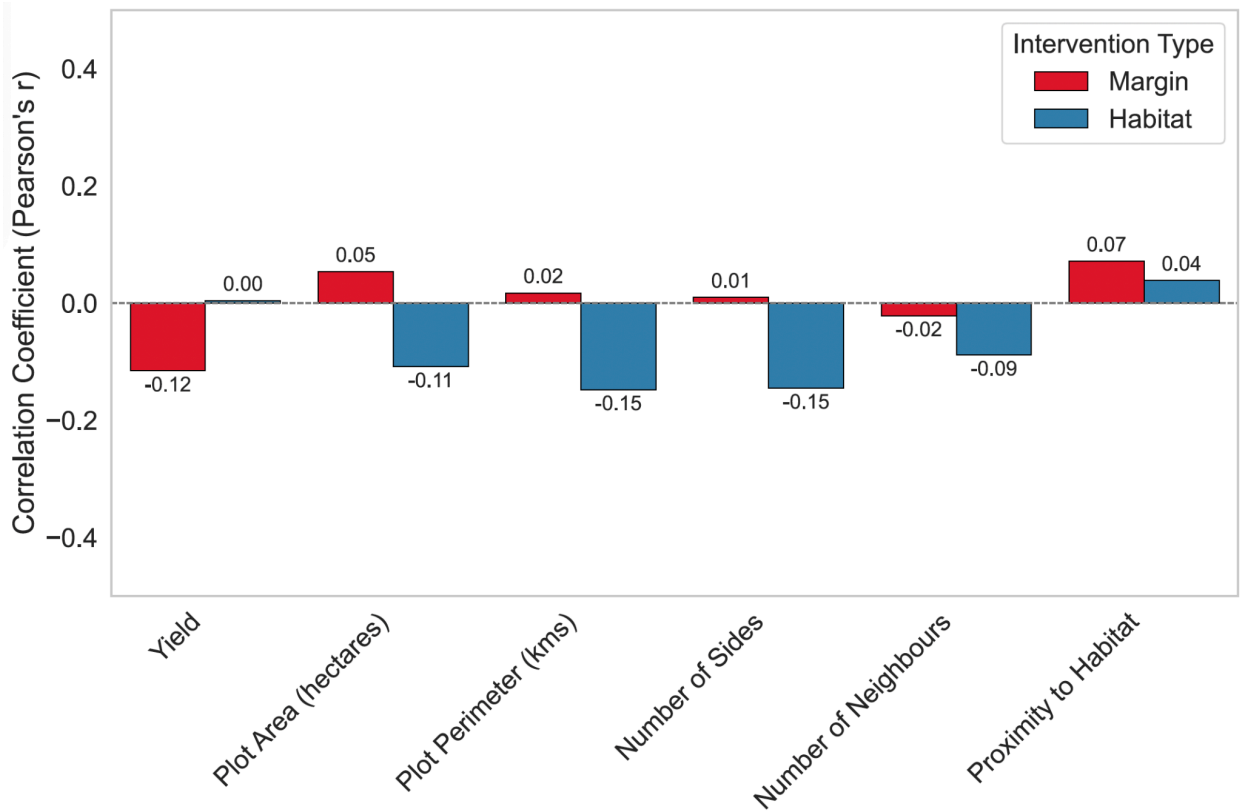

**Fig. 1: Pearson correlation coefficients between key plot factors and the optimal levels of margin and habitat interventions.** Shown for distance threshold of 1000 m and penalty coefficient of 1e3. The analysis revealed weak relationships between plot characteristics and the optimal intervention strategies selected by the model.

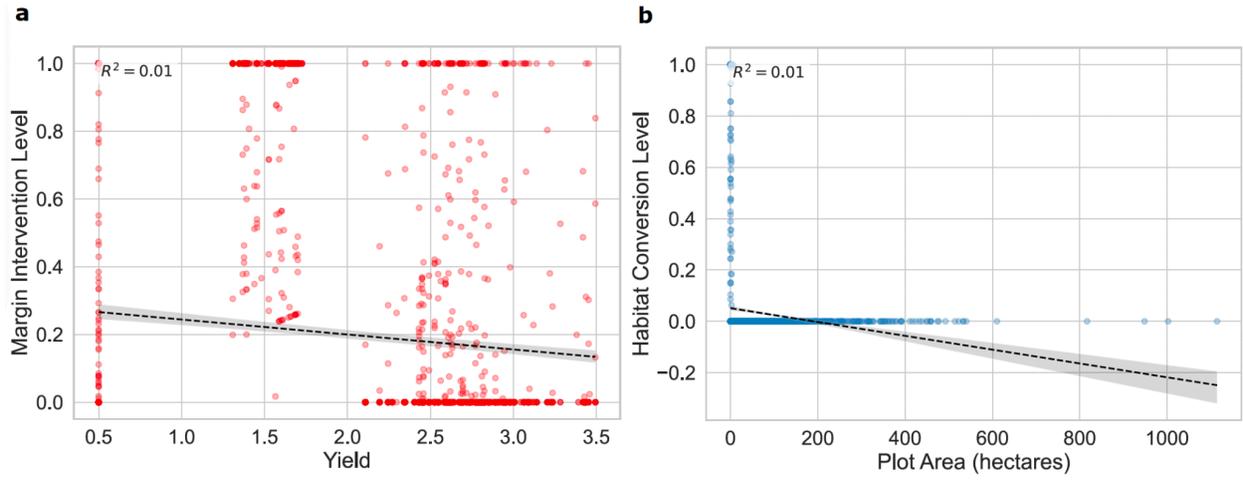

**Fig. 2: Scatter plots for intervention levels and selected parameters. a)** Scatter plot of margin intervention level and plot yield. The effect of yield appears to be not significant. **b)** Scatter plot of habitat conversion level and plot area. Smaller plots are strongly preferred for habitat conversion.

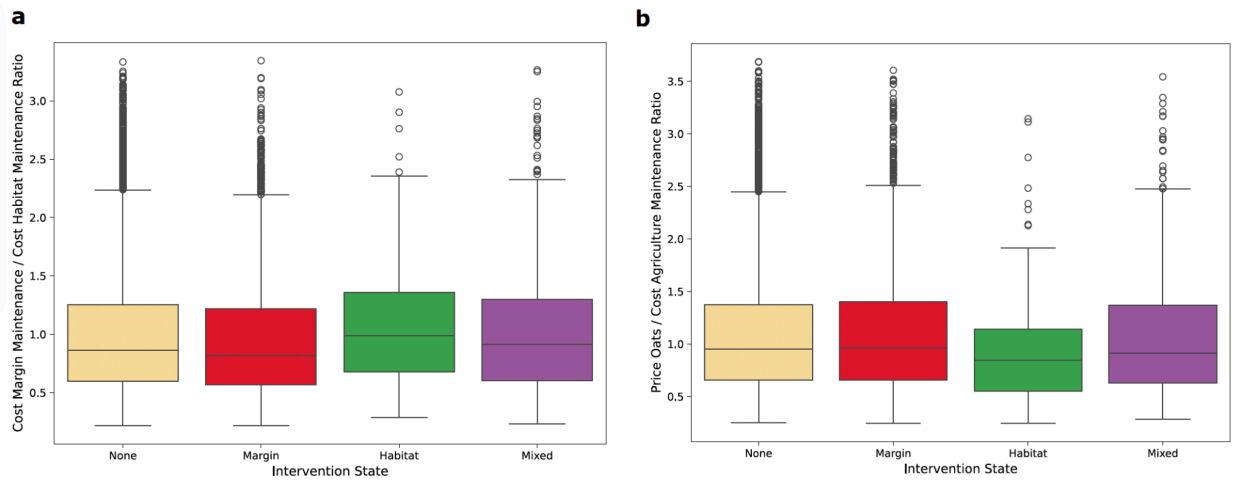

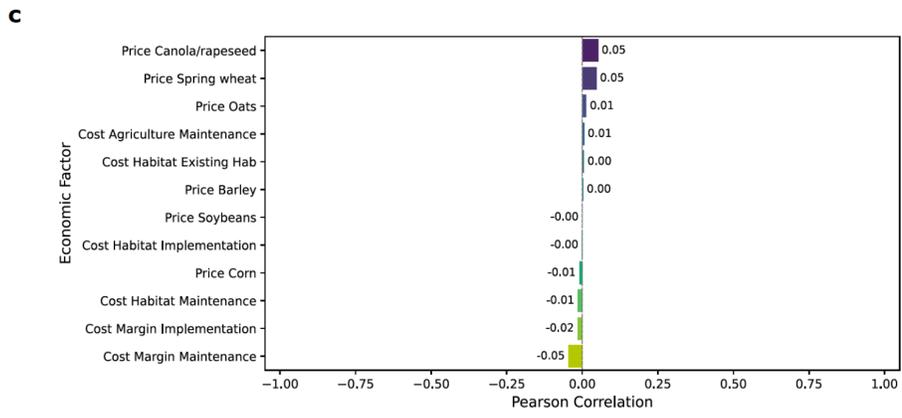

**Fig. 3: Sensitivity to costs and prices. a)** A box plot of intervention states as a function of the ratio of margin maintenance and habitat maintenance costs. **b)** A box plot of intervention states as a function of the ratio of price of Oats and agricultural maintenance costs. **c)** Sensitivity of margin intervention levels to crop prices and costs.

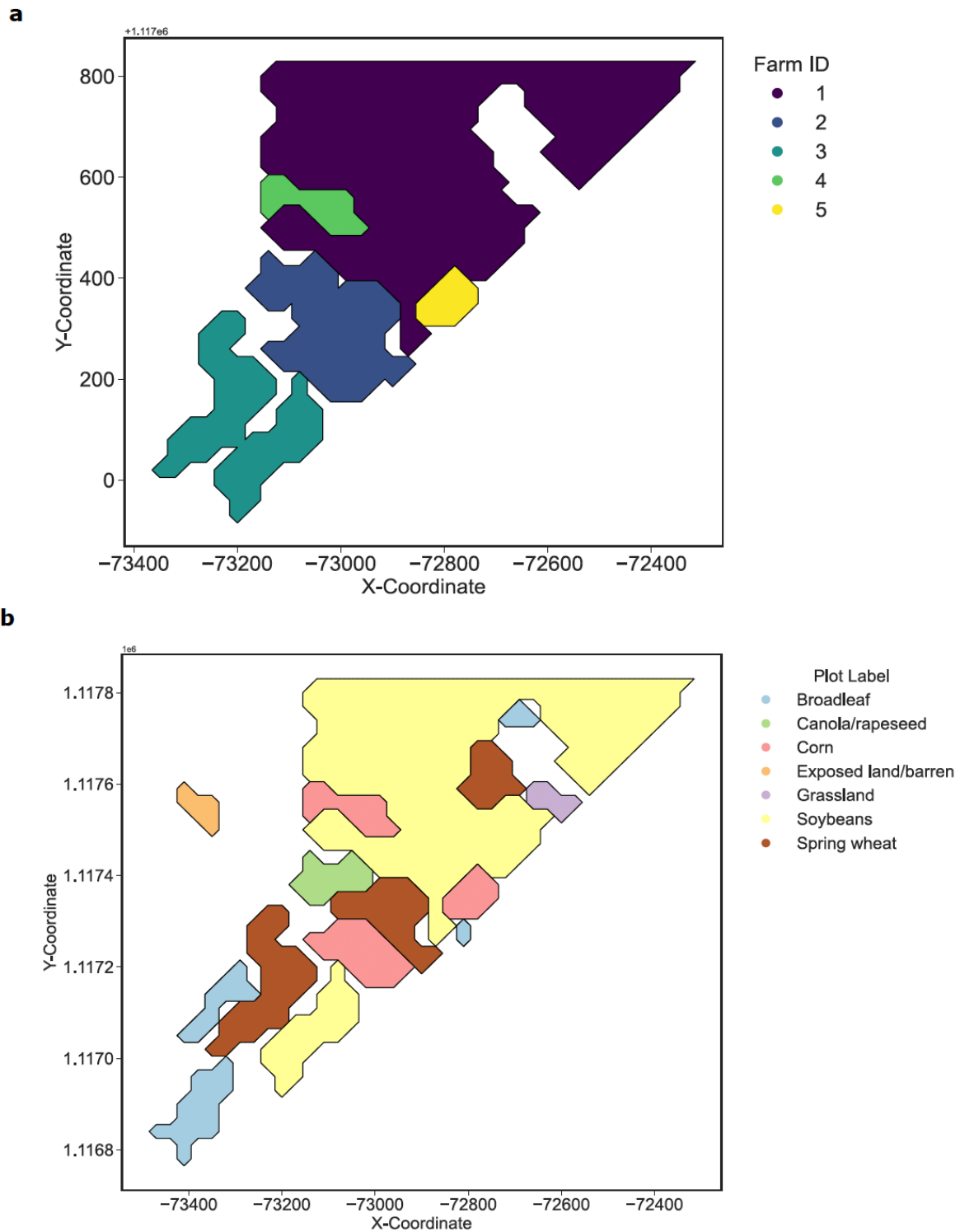

**Fig. 4: A selected configuration. a)** Farms in the configuration. **b)** Plots in the configuration.

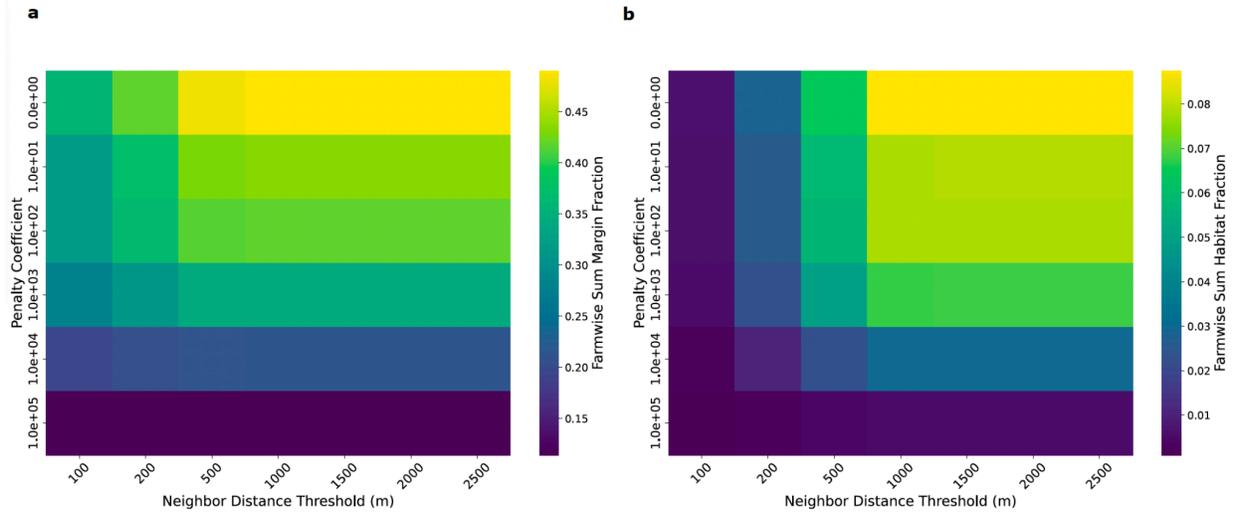

**Fig. 5: Heatmap of farmwise sum of intervention values with respect to the penalty coefficient and neighbour distance threshold. a)** For margin interventions. **b)** For habitat conversions. For small values of distance threshold, habitat allocation is extremely low. Therefore we set this threshold to 1000 m. We also observe that for small values of the penalty coefficient ($\lambda$) margin interventions always take extreme values of either 0 or 1. Therefore, we set the penalty coefficient to 1e3, to get a good spread of fractional interventions as well as achieve regularization.

A key strength of the model is its explicit incorporation of spatial, temporal, and crop-specific dynamics influencing the provision of pollination and pest control services generated by interventions. The benefits from new margins or habitats accrue over time, reflecting ecological processes like vegetation establishment or beneficial insect population growth. This is modeled via an exponential saturation function, $(1 - e^{-\gamma t})$, where the parameter γ (or ζ for pest control) dictates the rate of maturation. As illustrated in Fig. 6 and specifically for different crops in Fig. 7 (left panel), higher γ values lead to faster attainment of the maximum potential effect level. Ecosystem service effects diminish with distance from the source intervention. The model captures this using an exponential decay function, $e^{-\beta d_{ij}}$ (or $e^{-\epsilon d_{ij}}$ for pest control), where $d_{ij}$ is the distance between plot centroids and β (or ϵ) controls the decay rate. Plots receive benefits from their own interventions ($d_{ii} \approx 0$) and distance-discounted benefits from neighbors (within neighbor distance threshold). Fig. 6 shows the general sensitivity to the decay parameter, while Fig. 7 illustrates how this spatial reach varies depending on the crop, service, and intervention type, showing the effect level at a mature time point (T=50) across distances (right panel).

The model recognizes that different crops exhibit varying dependencies on pollination and benefit differently from pest control. Crucially, the parameters governing the magnitude, temporal dynamics, and spatial decay of these services are crop-specific. Fig. 7 explicitly visualizes this, demonstrating how the "Effect Level" for both pollination and pest control, generated by either margins or habitat conversion, varies significantly between crop groups (e.g., Canola/Rapeseed showing higher pollination benefits than Soybeans or Cereals).

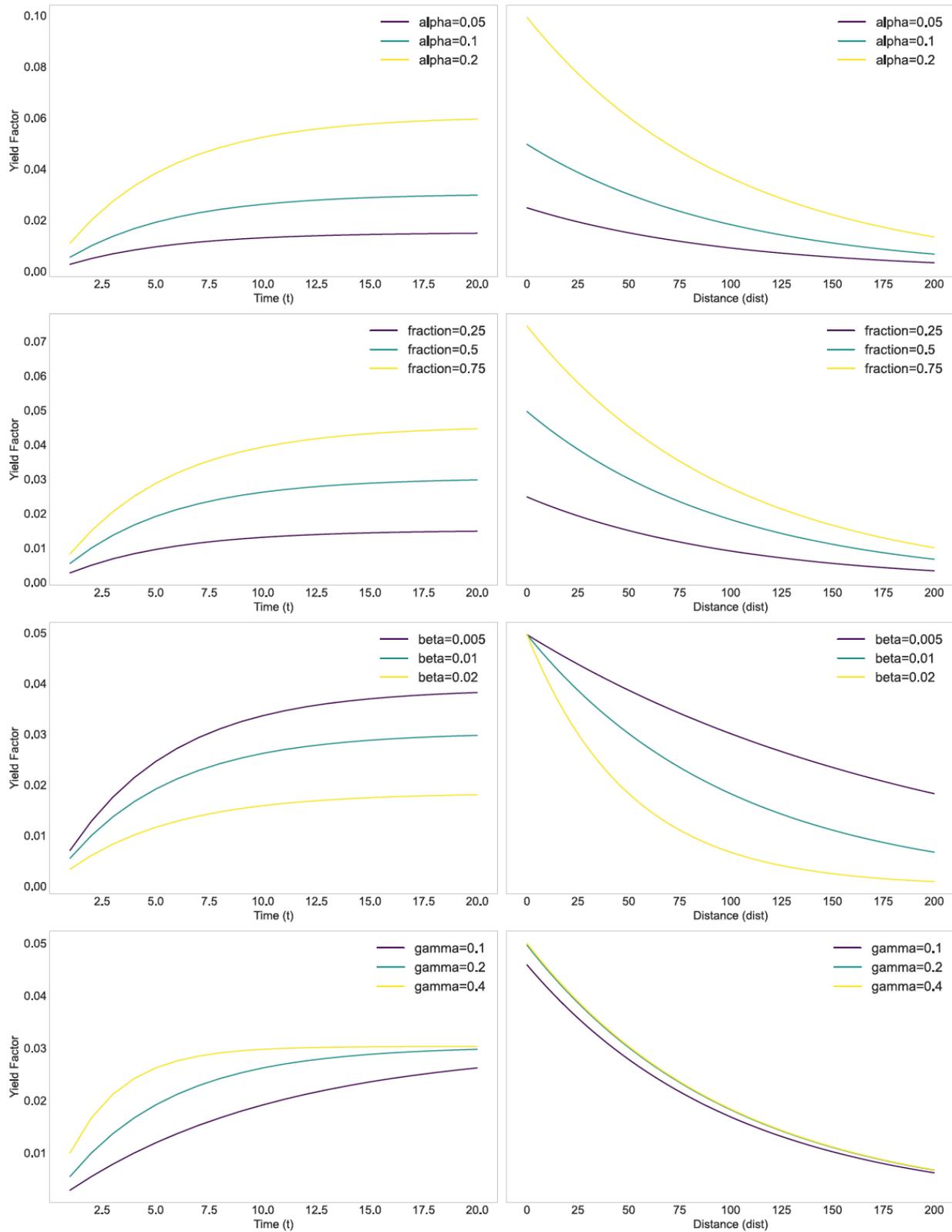

**Fig. 6: Yield factor as a function of parameters.** Yield factor is the multiplying factor that enhances yield as a result of ecosystem services. Left panel shows behavior with time, and right panel shows behavior with distance.

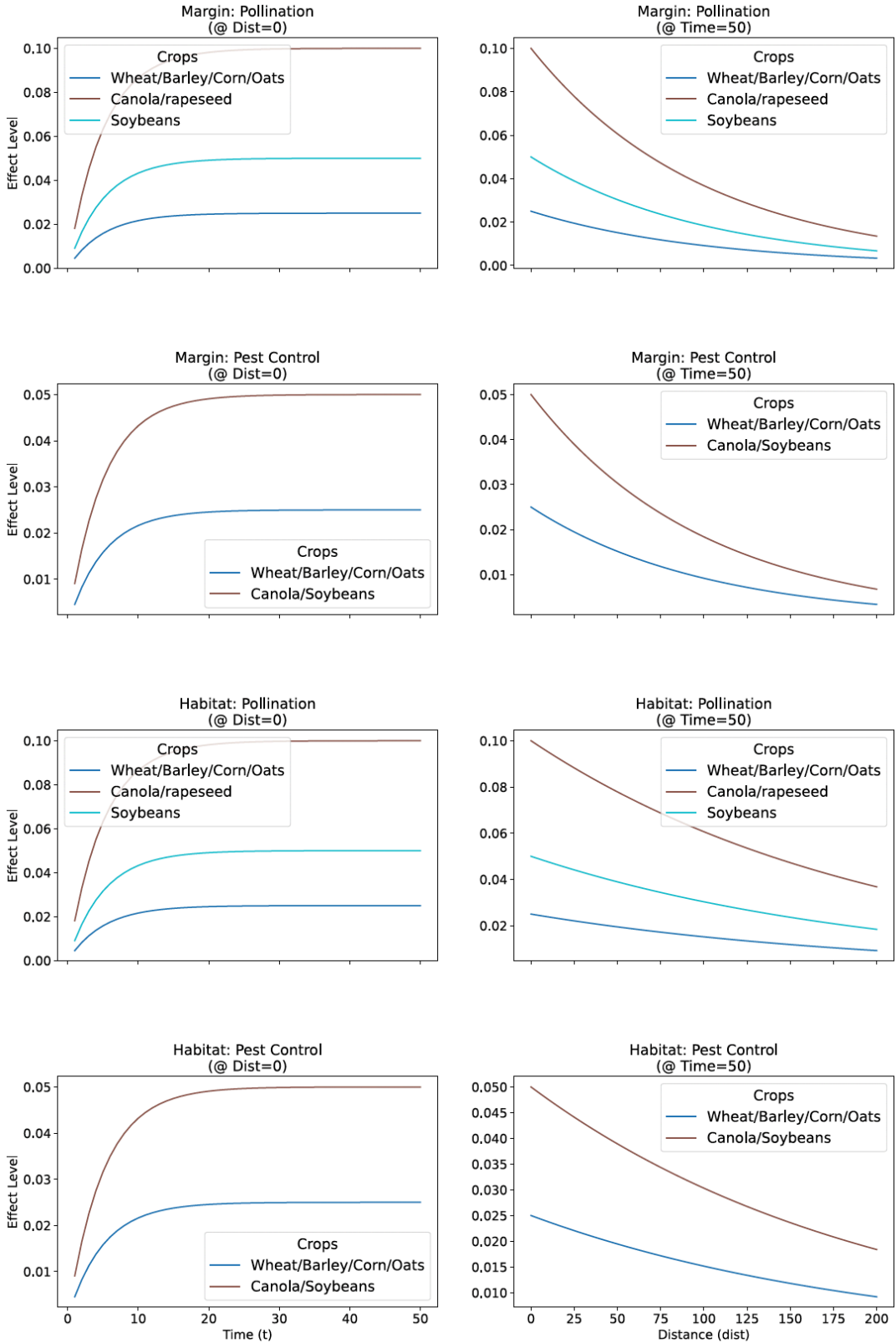

**Fig. 7: Effect level as a function of time and distance.** Crops are grouped if they share the same parameters. Left panel shows behavior with time, and right panel shows behavior with distance.

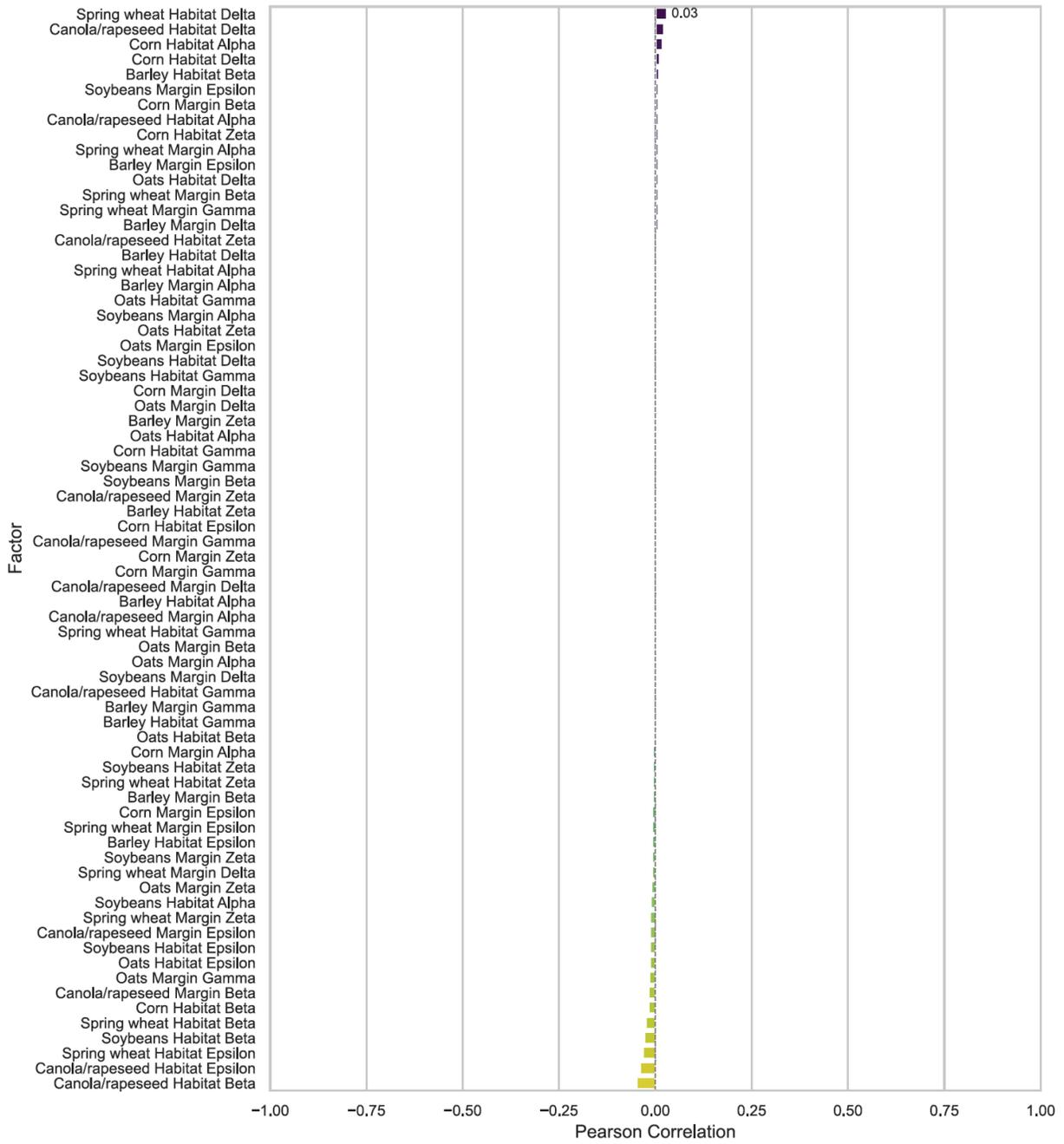

**Fig. 8: Sensitivity of habitat interventions to crop specific parameters.** Canola/rapeseed and spring wheat parameters show the highest sensitivity, with the rest being negligible.

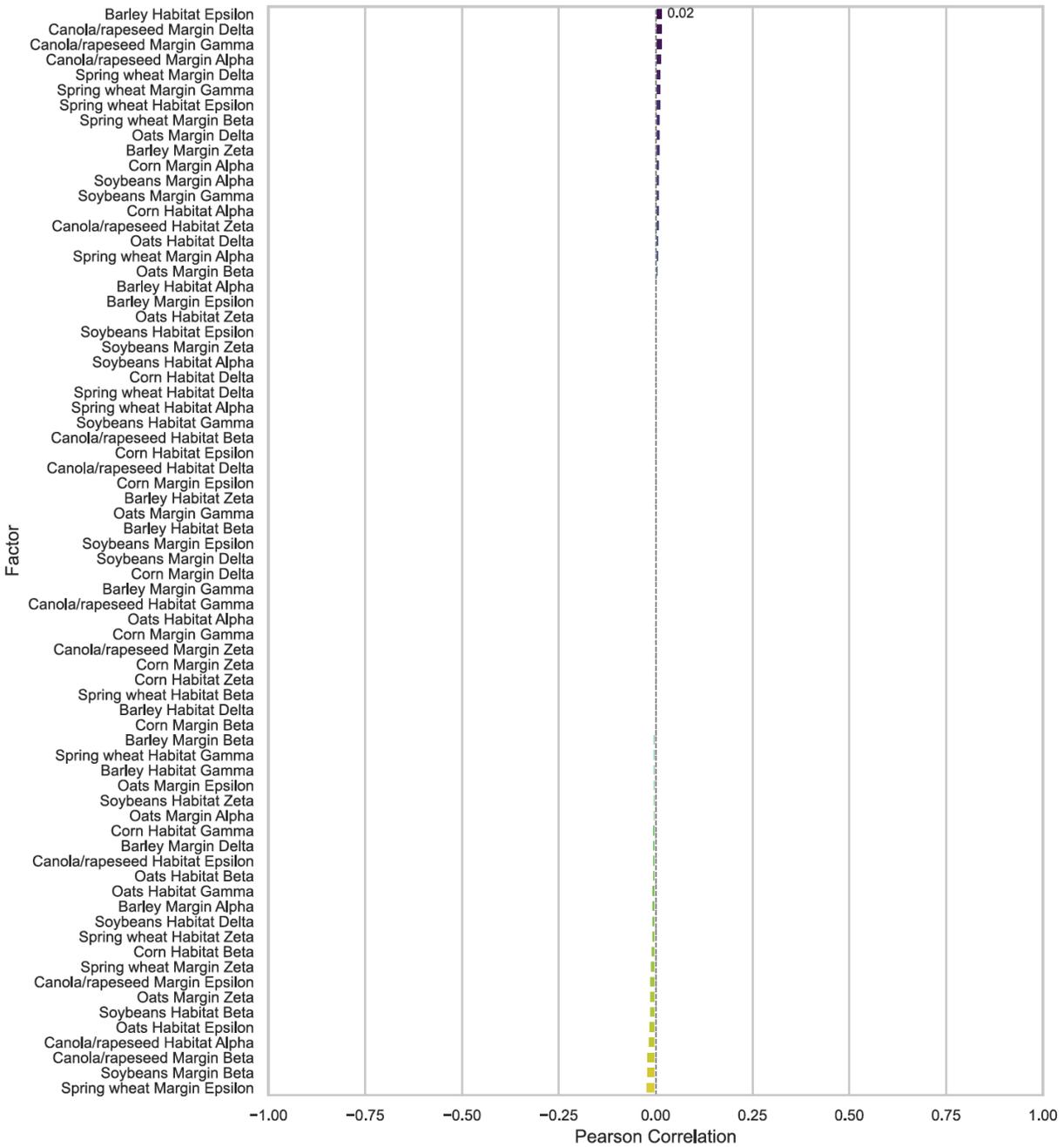

**Fig. 9: Sensitivity of margin interventions to crop specific parameters.** Most crop parameters have negligible effects on intervention fractions.

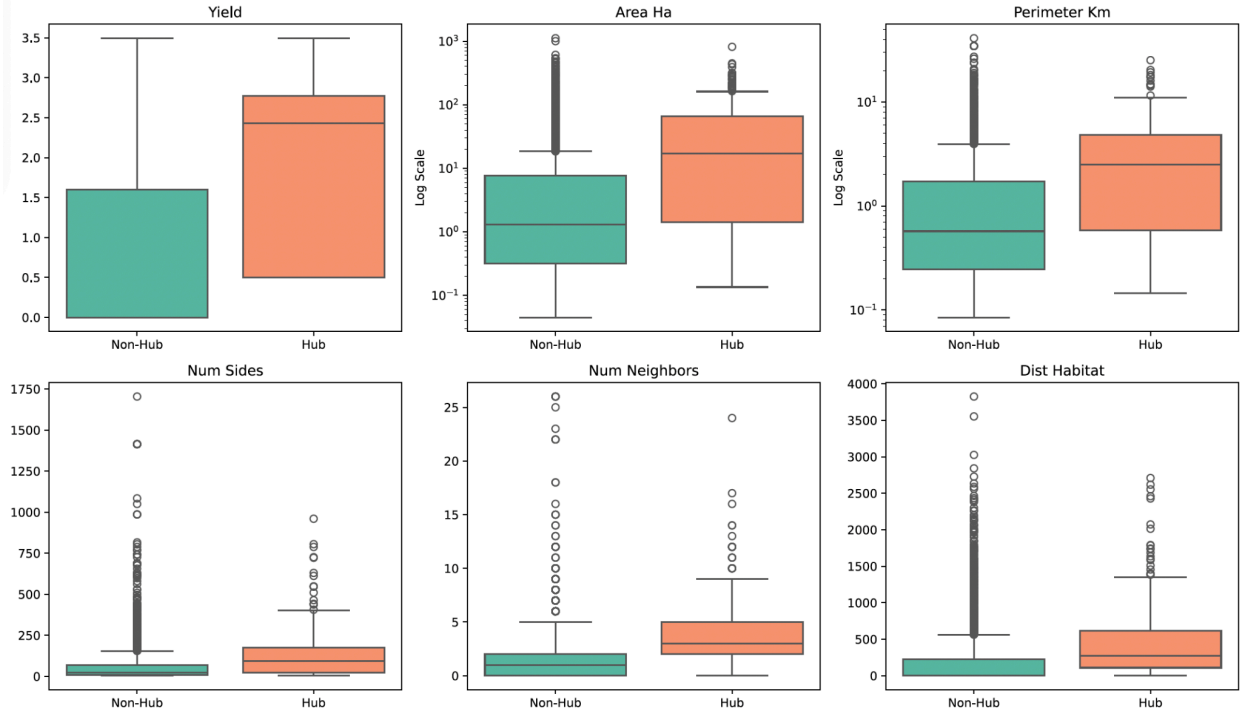

**Fig. 10: Box plots of aggregate parameters for hub and non-hub plots.** Showing baseline yield, area, perimeter, number of sides, number of neighbours, and distance to habitat.

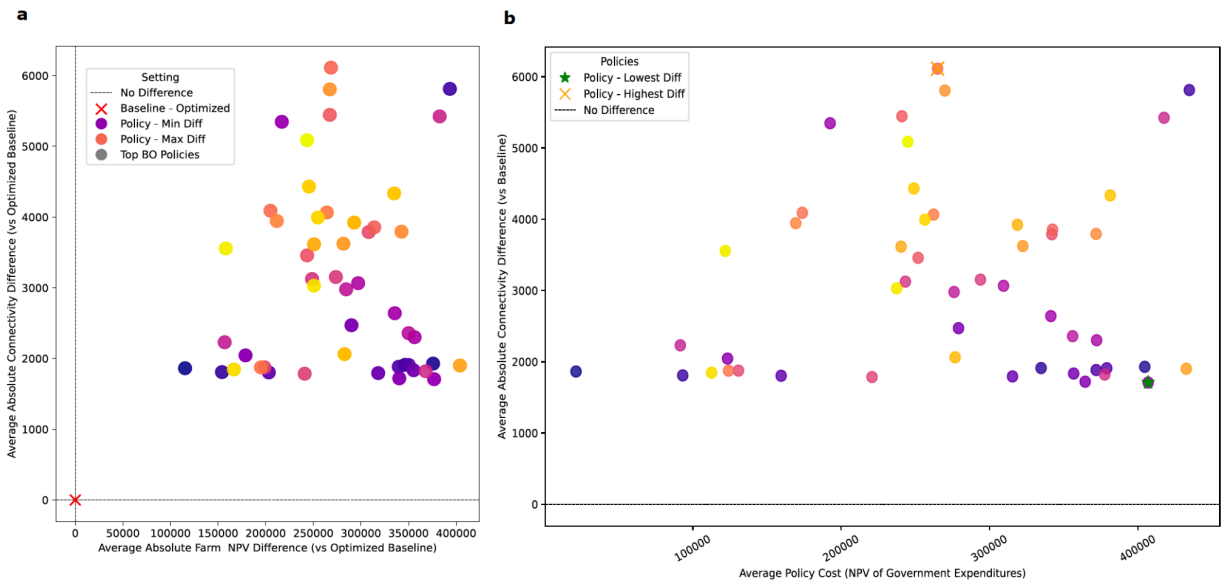

**Fig. 11: Scatter plots for top performing policies. a)** Average absolute connectivity difference with optimized baseline vs average absolute farm NPV difference with optimized baseline for top 50 policies identified by BO. **d)** Average absolute connectivity difference with optimized baseline vs average policy cost for top 50 policies identified by BO.

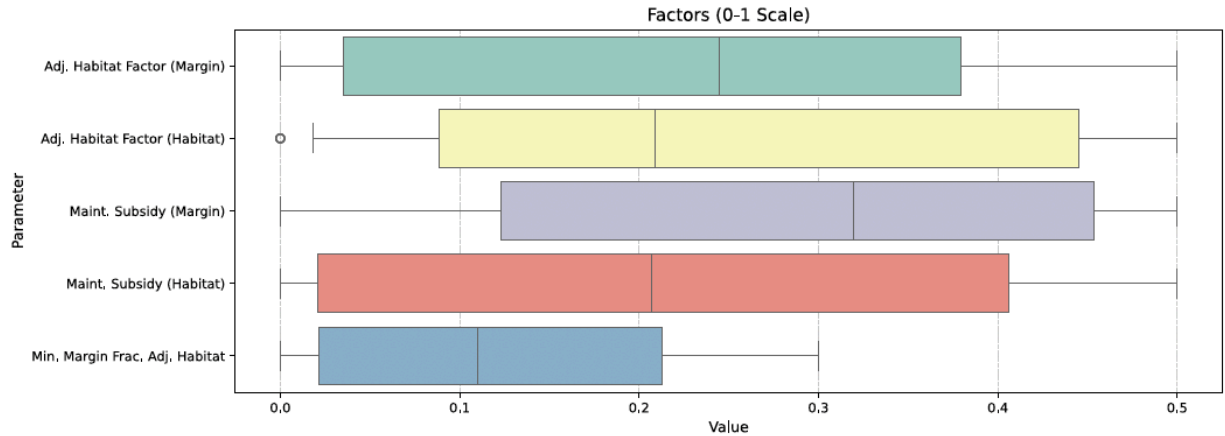
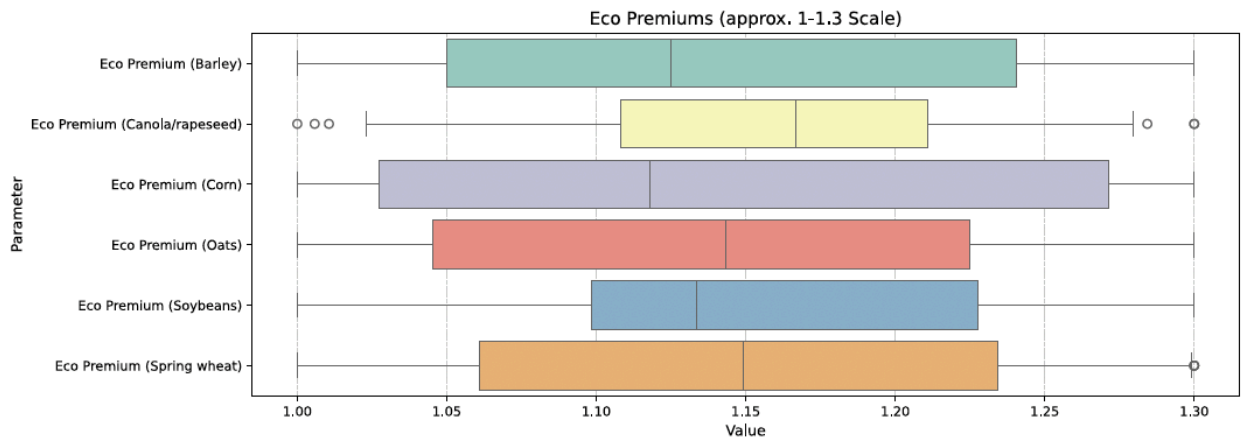
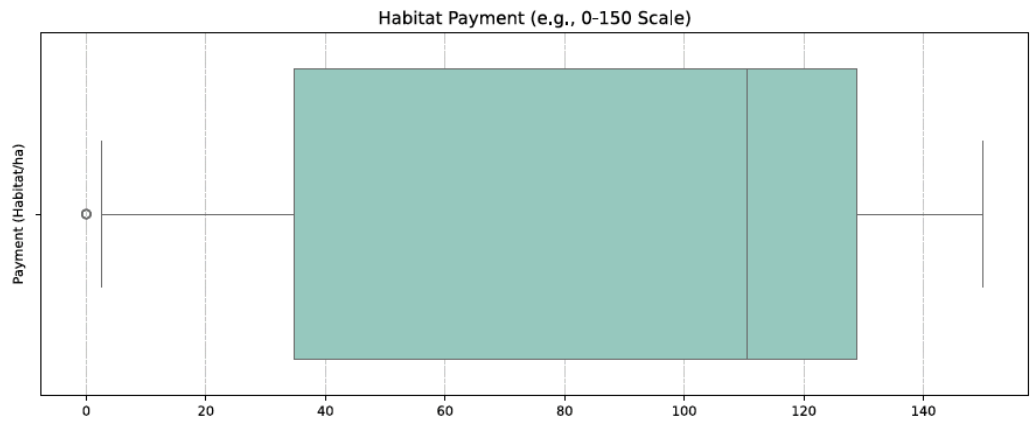
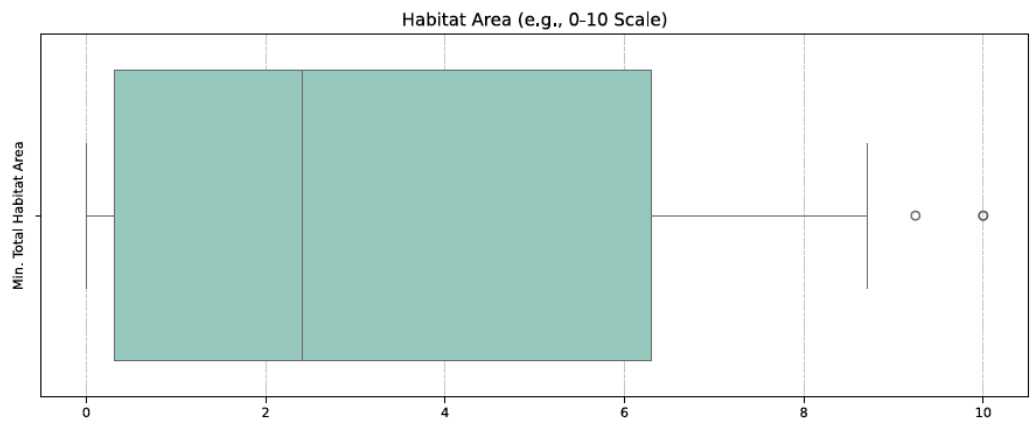

**Fig. 12: Key policy parameter distributions in top-performing policies.** Policies are grouped according to ranges and include subsidies, payments, mandates, and eco-premiums.

| Feature | Closest Conn (+-100), Max NPV & Closest Conn to Baseline & Min Conn Diff | Closest Conn (+-100), Min Cost | Min Conn Diff, Min Cost | Min Combined (Conn+NPV) Diff & Min NPV Diff |
|---|---|---|---|---|
| Original Policy ID | BO Policy 14 | BO Policy 12 | BO Policy 8 | BO Policy 1 |
| adj_hab_factor_margin | 0 | 0.0006692813581 | 0.1869007134 | 0.1407721897 |
| adj_hab_factor_habitat | 0.4986148311 | 0.02384206262 | 0.1172564376 | 0 |
| maint_subsidy_factor_margin | 0 | 0.08272880269 | 0.4939976433 | 0.5 |
| maint_subsidy_factor_habitat | 0.4225139696 | 0 | 0.3829979751 | 0.003652709541 |
| hab_per_ha | 112.8883511 | 147.8696113 | 116.5506658 | 145.317865 |
| min_total_hab_area | 0.5554092517 | 0.4937111504 | 0.2798195504 | 0 |
| min_margin_frac_adj_hab | 0.2746218434 | 0.2891910735 | 0.05217195487 | 0 |
| eco_premium_factor_Barley | 1.082518439 | 1.036696299 | 1.046224673 | 1.261281479 |
| eco_premium_factor_Canola/rapeseed | 1.244266555 | 1.162771886 | 1.023125944 | 1 |
| eco_premium_factor_Corn | 1.03586856 | 1.3 | 1.26695971 | 1 |
| eco_premium_factor_Oats | 1.222501986 | 1.010158569 | 1.225113611 | 1 |
| eco_premium_factor_Soybeans | 1 | 1.3 | 1.208020973 | 1 |
| eco_premium_factor_Spring wheat | 1.3 | 1.274470364 | 1.153529013 | 1.003099153 |
| avg_conn | 7229.614074 | 7261.595073 | 7035.698245 | 6820.273505 |
| avg_policy_cost | 406972.9037 | 364486.2791 | 159560.4806 | 21254.18969 |
| avg_farm_npv | 627763.1374 | 596348.6474 | 475336.8733 | 397510.8454 |
| avg_conn_diff | 1708.409763 | 1720.272549 | 1803.648634 | 1863.149882 |
| avg_npv_diff | 376873.9032 | 340285.2972 | 203275.5768 | 115288.468 |
| conn_abs_diff_from_baseline | 19.52332885 | 51.50432822 | 174.3925002 | 389.8172403 |

**Table 1: Policy instrument parameters for specific successful policy archetype.** These include closest connectivity to baseline, closest connectivity (+-100) and maximum NPV, minimum connectivity difference with baseline per farm, closest connectivity (+-100) and minimum cost, minimum connectivity difference with baseline per farm (+-100) and minimum cost, minimum combined (connectivity+NPV) difference with baseline per farm.

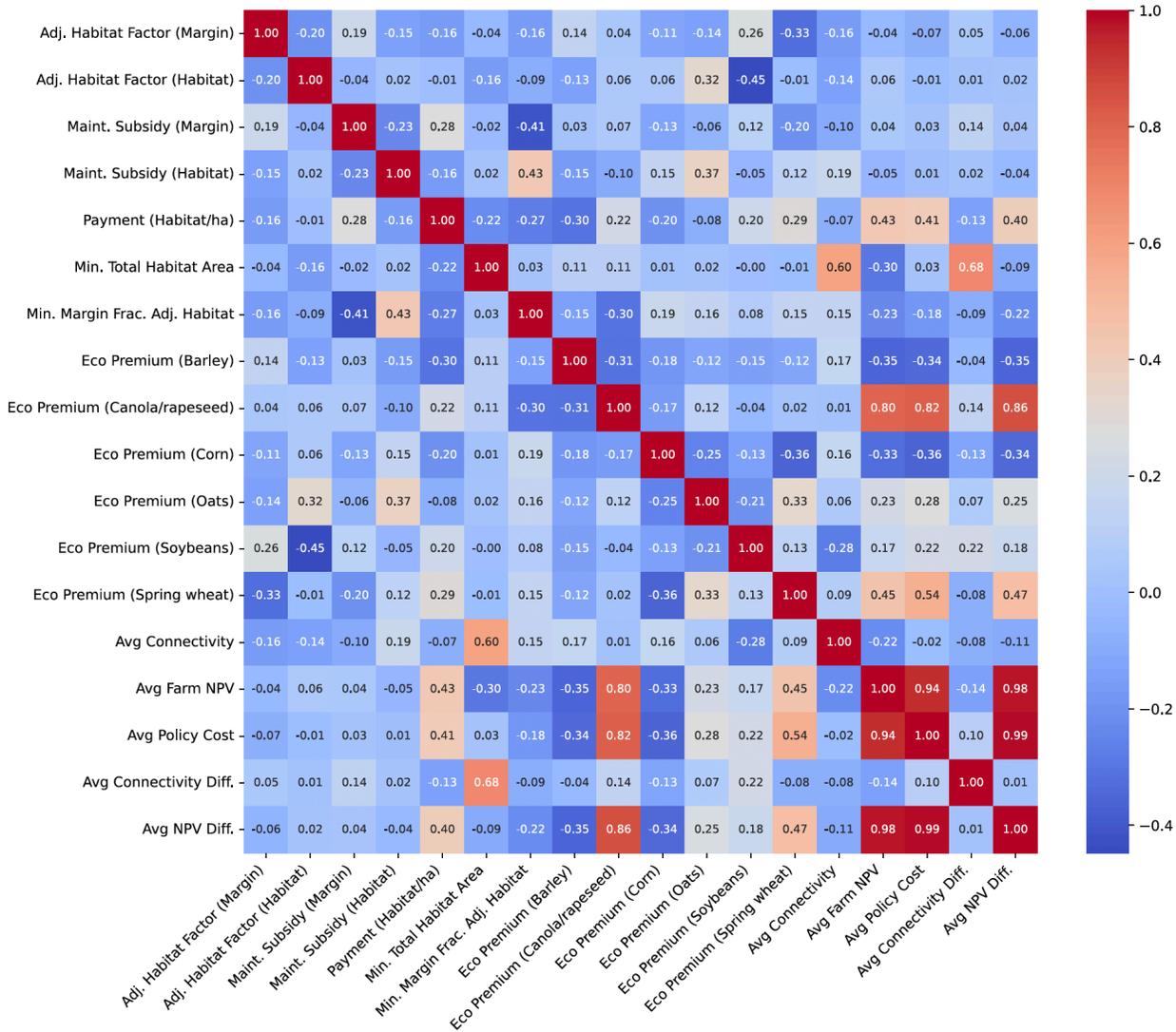

**Fig. 13: Correlation matrix of policy parameters and policy metrics.** Policy parameters include subsidies, payments, mandates, and eco-premiums, and metrics include connectivity, NPV, policy cost, farm-wise connectivity difference and NPV difference with optimized baseline.

## Synthetic Data Experiments

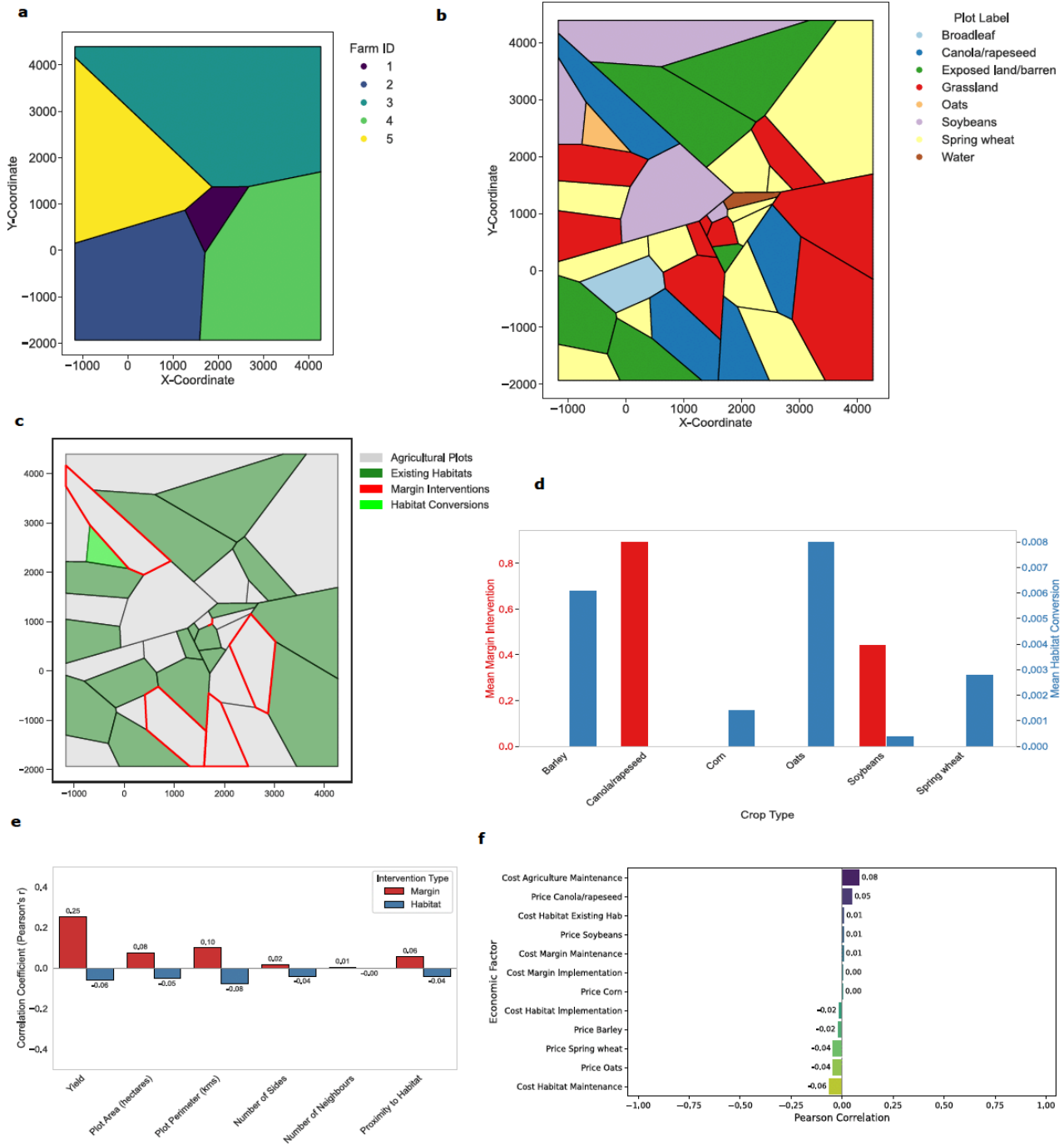

**Fig. 14: EI model results on synthetic data. a)** Generated synthetic farm-level geospatial data for one of the landscape configurations. The configuration consists of a total of five farms with five-ten plots per farm. See the "Methods" section for details on synthetic data creation. **b)** Generated synthetic plot-level geospatial data for one of the landscape configurations. The configuration consists of a total of five farms with five-ten plots per farm. **c)** Optimal margin and habitat fractions decided by the EI model shown for farms. The spatial location of fractions is chosen at random for visualization. **d)** Margin and habitat intervention fractions chosen based on crop type. **e)** Pearson correlation coefficients between key plot factors and the optimal levels of margin and habitat interventions. **f)** Sensitivity of habitat conversion levels to crop prices and costs.

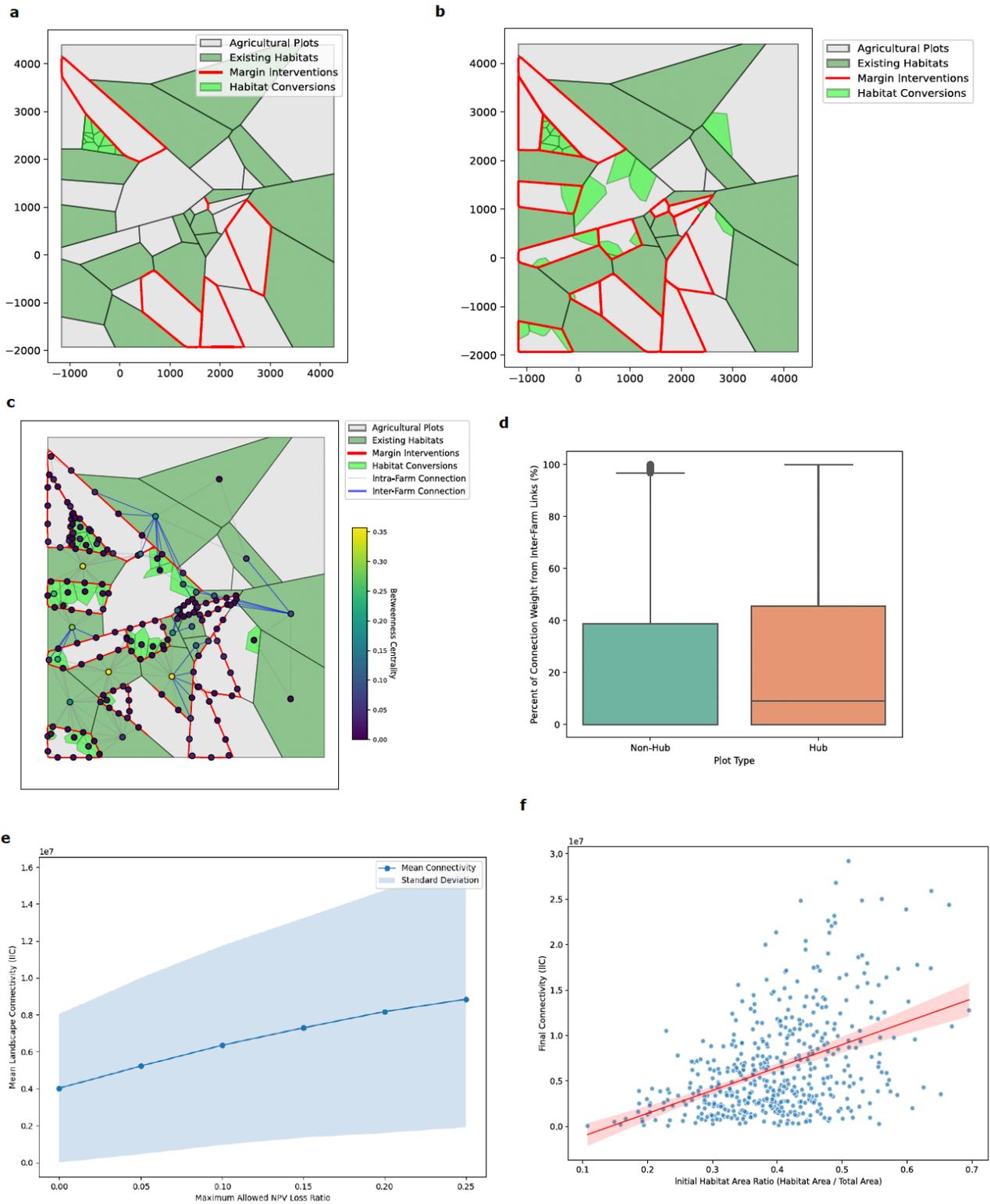

**Fig. 15: EC model results on synthetic data. a)** Margin arcs and habitats areas chosen after EC repositioning. **b)** Margin arcs and habitats areas chosen after EC connectivity optimization. **c)** Plot level betweenness centrality and intra-farm (dashed gray)/inter-farm (solid blue) connections. **d)** Box plot of aggregate yield for hub and non-hub plots. **e)** Mean landscape

connectivity (IIC) vs maximum allowed NPV loss ratio, with mean and standard deviation shown. **f)** Scatter plot of optimized connectivity (IIC) vs initial existing habitat area ratio.

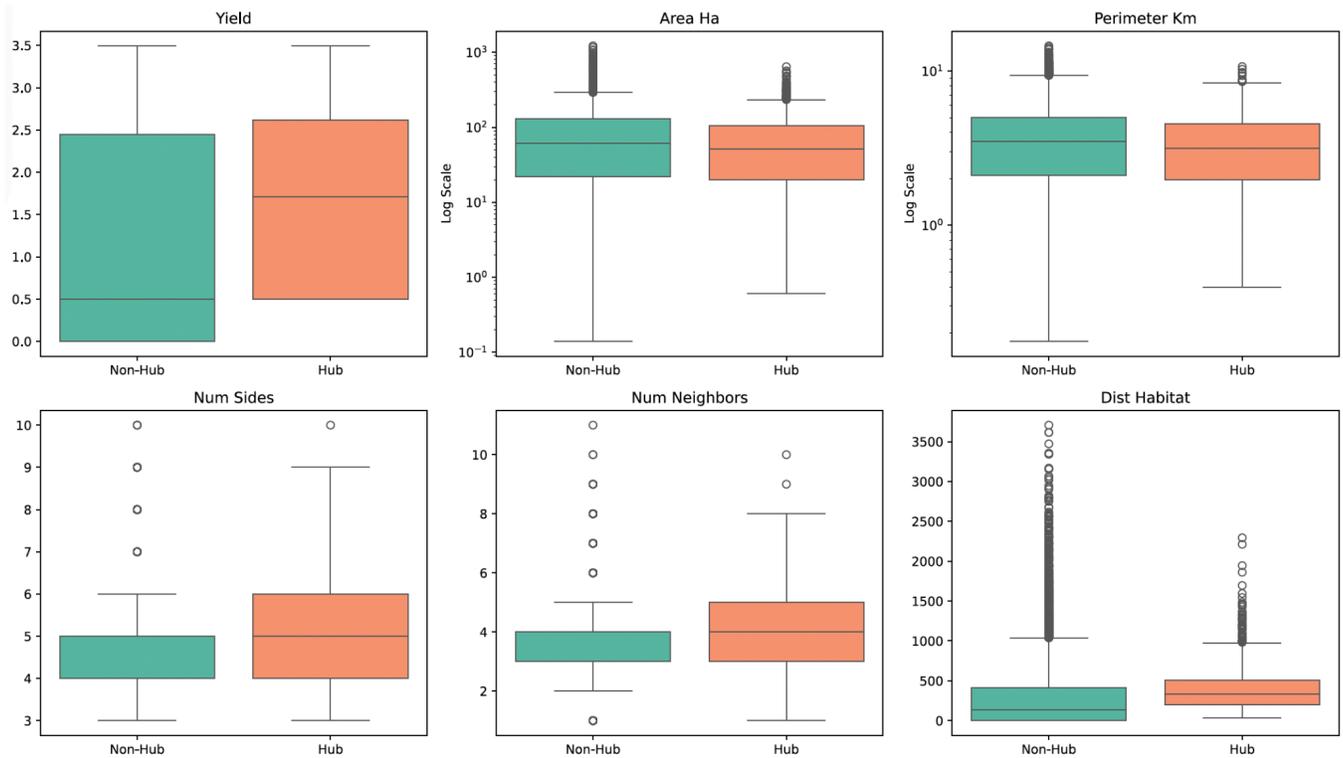

**Fig. 16: Box plots of aggregate parameters for hub and non-hub plots.** Showing baseline yield, area, perimeter, number of sides, number of neighbours, and distance to habitat.

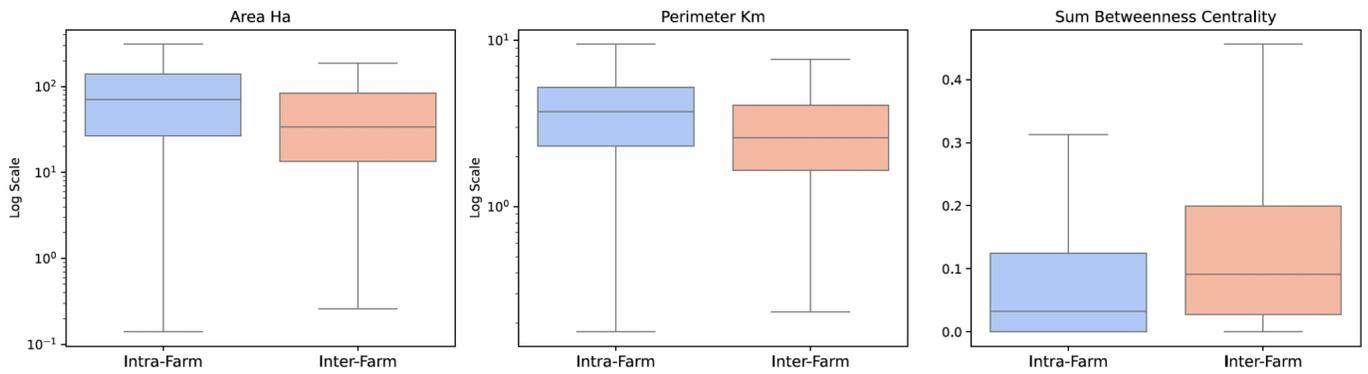

**Fig. 17: Box plots of aggregate parameters for plots with intra-farm and inter-farm dominating connections.** Showing area, perimeter, and sum betweenness centrality.